\documentclass[amsmath,prl,twocolumn,showpacs,superscriptaddress]{revtex4}

\usepackage{amssymb}
\usepackage{graphicx}

\begin{document}

\title{Observation and analysis of Fano-like lineshapes in the Raman spectra of molecules adsorbed at metal interfaces}

\author{S. Dey}
\affiliation{Department of Chemistry, University of California, Irvine, California 92697-2025, USA}
\author{M. Banik}
\affiliation{Department of Chemistry, University of California, Irvine, California 92697-2025, USA}
\author{E. Hulkko}
\affiliation{Department of Chemistry, University of California, Irvine, California 92697-2025, USA}
\affiliation{Depratment of Chemistry, PO Box 35, FI-40014, University of Jyv\"askyl\"a, Finland}
\author{K. Rodriguez}
\affiliation{Department of Chemistry, University of California, Irvine, California 92697-2025, USA}
\author{V. A. Apkarian}
\affiliation{Department of Chemistry, University of California, Irvine, California 92697-2025, USA}
\author{M. Galperin}
\affiliation{Department of Chemistry and Biochemistry, University of California at San Diego, La Jolla, CA 92093, USA}
\author{A. Nitzan}
\affiliation{Department of Chemistry, University of Pennsylvania, Philadelphia, Pennsylvania 19104, USA}
\affiliation{School of Chemistry, Tel Aviv University, Tel Aviv, 69978, Israel}

\begin{abstract}
Surface enhanced Raman spectra from molecules (bipyridyl ethylene) adsorbed on gold dumbells 
are observed to become increasingly asymmetric (Fano-like) at higher incident light intensity. 
The electronic temperature (inferred from the anti-Stokes (AS) electronic Raman signal 
increases at the same time while no vibrational AS scattering is seen. 
These observations are analyzed by assuming that the molecule-metal coupling 
contains an intensity dependent contribution (resulting from light-induced charge transfer 
transitions as well as renormalization of the molecule metal tunneling barrier). 
We find that interference between vibrational and electronic inelastic scattering routes 
is possible in the presence of strong enough electron-vibrational coupling and can 
in principle lead to the observed Fano-like feature in the Raman scattering profile. 
However the best fit to the observed results, including 
the dependence on incident light intensity and the associated thermal response is obtained 
from a model that disregards this coupling and accounts for the structure of the continuous 
electronic component of the Raman scattering signal. The temperatures inferred from the Raman 
signal are argued to be only of qualitative value. 
\end{abstract}


\maketitle


 Surface enhanced optical response of molecules adsorbed on metal surfaces 
reflects the effect of strong local fields created by 
surface plasmons\cite{MoskovitsRMP85,GerstenNitzanSurfSci85,GerstenPlasmonics07}
 as well as charge transfer between the molecule and metal.\cite{GerstenBirkeLombardiPRL79,OttoJPCM92,OttoFutamata,JensenCR11}
Surface enhanced Raman scattering (SERS) has become an important diagnostic tool for molecules 
at metallic interfaces including molecular conduction junctions.\cite{HoPRB08,CheshnovskySelzerNatNano08,NatelsonNL08,TianNatComm11,SelzerChemSocRev11,NatelsonNatNano11,KiguchiMurakoshiJACS13,NatelsonPCCP13} (See also Refs.~\onlinecite{MGANPCCP12,GalperinRatnerNitzanNL09,GalperinRatnerNitzanJCP09,ParkMGEPL11,ParkMGPRB11,RatnerJPCC13,OrenMGANPRB12,ParkMGPST12,KaasbjergNitzanPRB13,WhiteTretiakNL14} 
for related theoretical work). At metal interfaces, vibrational Raman scattering is accompanied by 
a continuous background resulting from inelastic contributions of electron-hole (e-h) excitations in 
the metal.\cite{MoskovitsRMP85,GerstenBirkeLombardiPRL79,BursteinSolStCommun79,AkemannOttoSurfSci94,ItohJCP06} 
Both components of the inelastic scattering signal were recently used to determine the bias induced 
heating in a molecular conduction junction.\cite{NatelsonNatNano11}

 While it is generally recognized that the electronic background is affected by 
 the molecule-substrate interaction, these two components of the SERS signal are usually treated 
 separately. Surprizingly, we observe (Fig.~\ref{fig1}) an apparent Fano-like 
 feature that may indicate interference between these two scattering channels.
Below we discuss the origin of this observation. 

\begin{figure}[htbp]
\centering\includegraphics[width=0.9\linewidth]{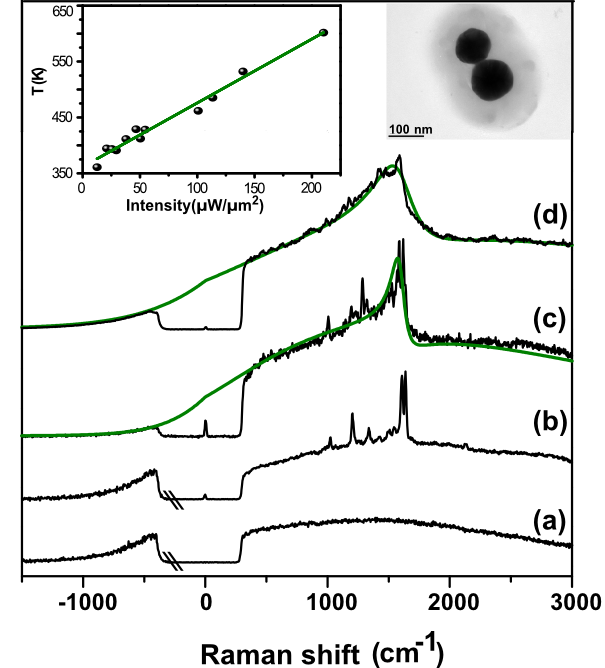}
\caption{\label{fig1}
Raman spectrum recorded on 
(a) a bare dumbbell, without reporter molecules at low irradiation intensity 
($\sim 10\ \mu$W/$\mu $m${}^{2}$); 
(b) with reporter molecules at the same low intensity; 
(c) and (d) with reporter molecules at irradiation intensities $\sim 50\ \mu$W/$\mu$m${}^{2}$ 
and $\sim 150\ \mu$W/$\mu $m${}^{2}$, respectively. 
Demarcations on (a) and (b) indicate a magnification of the anti-Stokes (AS) region (7x and 4.5x, 
respectively). The green traces overlapping the (c) and (d) lines are fits to the Fano lineshape (see SI). 
The right inset shows a TEM image of a typical dumbbell. The left inset shows the temperature 
extracted from the fit of the AS branch of the electronic Raman spectrum (continuum) to the 
Fermi Dirac distribution. No vibrational AS signal is detected.
}
\end{figure}

The measurements are carried out on single silica encapsulated gold dumbbells 
(Fig.~\ref{fig1} right inset). 
The nanosphere diameter is $95\pm 5$ nm and the intersphere spacing prior to irradiation is 
$\sim 1$~nm.  
On these dumbbells, the quadrupolar and the binding dipolar plasmon resonances occur near 
$560$~nm and $780$~nm, respectively.\cite{MarhabaJPCC09}
As molecular reporter, bipyridyl ethylene (BPE) is adsorbed on the gold 
spheres prior to encapsulation. The dumbbells are dispersed on a silicon nitride membrane (20 nm 
thick) of the TEM grid by drop casting them in a dilute solution. Locations and 
geometries of the nanostructures are mapped 
using a scanning electron microscope (SEM), then Raman scattering 
measurements are done under an optical microscope, in the backscattering geometry, using an 
objective with a numerical aperture of NA${}=0.625$. The excitation source is a continuous wave diode 
laser, operating at $\lambda=532$~nm, coincident with the anti-bonding quadrupolar surface 
plasmon resonance.\cite{MarhabaJPCC09} The molecular 
vibrational Raman spectra appear over a background continuum (Fig.~\ref{fig1}b), 
which is also seen on bare dumbbells (Fig.~\ref{fig1}a). 
As discussed before,\cite{ApkarianNatPhoton14} at low intensity, 
$10\ \mu$W/$\mu$m${}^{2}$, the vibrational spectrum 
matches that of the isolated BPE molecule.  
Two observations are most significant at higher incident intensity: 
(a) The molecular lines broaden asymmetrically to eventually coalesce into 
the single asymmetric profile, similar to Fano-type lineshape (Figs.~\ref{fig1}c and d). 
(b) The electronic and vibrational temperatures, $T_e$ and $T_v$, 
inferred from the AS branch of the electronic Raman scattering
and S/AS intensity ratio of the vibrational Raman line appear to be different. 
At the highest incident light intensity $T_e$ reached $580$~K, while no molecular 
AS scattering is seen, which would set the apparent vibrational temperature 
of the molecule, $T_v$, to less than $300$~K. More experimental details and results 
are presented in the SI. 

The central question raised by these observations is whether the lineshapes shown 
in Figs.~\ref{fig1}c, d are indeed interference features as their Fano-like lineshapes suggest. Specifically: 
can there be interference between the electronic and vibrational Raman scattering pathways 
that coexist in this system? 
The intuitive answer is negative, because the two pathways lead to different final states.
The model calculations presented below yield two main results:
(a) The aforementioned interference can exist and can in principle lead to
the observed lineshapes. 
(b) However, comparing the detailed calculations with the available experimental observations
suggests that another mechanism, asymmetric electronic sidebands dressing the molecular 
Raman spectrum, may be dominant. 

\begin{figure}[htbp]
\centering{\includegraphics[width=0.4\linewidth]{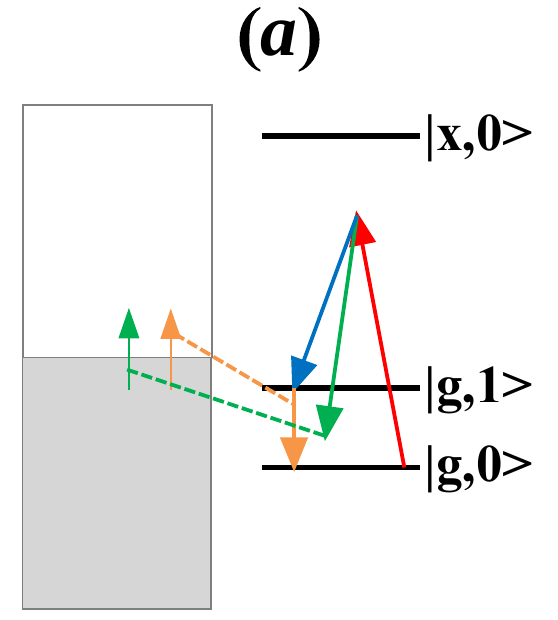}\ \ \ \ \ \ \includegraphics[width=0.4\linewidth]{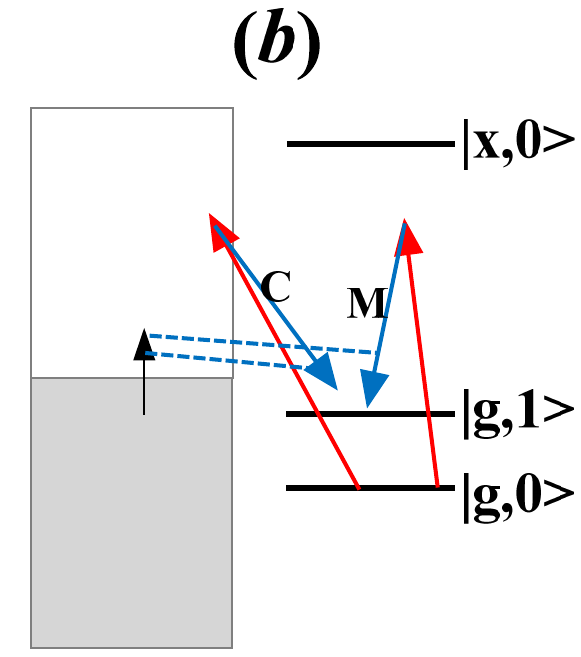}}
\caption{\label{fig2}
(Color online) A sketch of the processes underlying the Stokes (S) component of Raman scattering from 
a molecule coupled to a metal. (a) The incident photons induce the virtual transition displayed in red 
and a corresponding Rayleigh scattering (the red process in reverse). 
In addition two inelastic scattering (Raman) 
pathways are shown: The green process brings the molecule back to its ground state while creating an 
e-h excitation in the metal (electronic Raman scattering). The blue process brings the molecule to an 
excited vibrational state on the electronic ground state (vibrational Raman scattering). 
These Raman scattering pathways are coupled by the combined effect of the metal-molecule electron-
transfer coupling and the electronic-vibrational (e-v) interaction (orange) that induces vibrational de-
excitation in the molecule with e-h formation in the metal. The dashed lines connect coupled processes. 
(b) Two pathways are defined by the excitation process (red): A (virtual) molecular transition and a 
molecule-to-metal charge transfer transition, respectively denoted M and C in the text. The outgoing 
photon (blue) leaves the molecule in the excited vibrational state, possibly with e-h excitation (or de-
excitation) in the metal (black). Note that the processes shown are accompanied also by
electronic Raman scattering from the bare metal.
}
\end{figure}

We consider a molecule chemisorbed on a metal surface and subjected to an external radiation field. 
In the experimental system the metal is a nanostructure (NS) that promotes plasmonic 
enhancement, but this enters our discussion only implicitly, through the magnitude of the local 
electromagnetic field. The model Hamiltonian comprises the molecule, metal, radiation field and their 
couplings, with the electron-photon and e-v interaction terms treated as perturbations:
\begin{equation} \label{eq1} 
\hat{H}=\hat{H}_{0} +\hat{V}_{rad} +\hat{V}_{e-v}  
\end{equation} 
where ($\hbar=k_B=e=1$)
\begin{align} 
\label{H0} 
&\hat{H}_{0} = \hat{H}_{M} +\hat{H}_{NS} +\hat{V}_{NS} +\hat{H}_{rad} ;       
\\
\label{HM} 
&\hat{H}_{M} = \sum _{m=g,x} \varepsilon _{m} \hat{d}_{m}^{\dag } \hat{d}_{m} +\omega _{v} \hat{v}^{\dag } \hat{v};       
\\
\label{HNS} 
&\hat{H}_{NS} = \sum _{k} \varepsilon _{k} \hat{c}_{k}^{\dag } \hat{c}_{k}; 
\\
\label{VNS}
&\hat{V}_{NS} = \sum _{m=g,x} \sum _{k} \left(V_{mk} \hat{d}_{m}^{\dag } \hat{c}_{k} +V_{km} \hat{c}_{k}^{\dag } \hat{d}_{m} \right);
\\
 \label{Vev} 
&\hat{V}_{e-v} = \sum _{m=g,x} M_{m} \hat{Q}_{v} {\kern 1pt} \hat{d}_{m}^{\dag } \hat{d}_{m} ;  \hat{Q}_{v} =\hat{v}+\hat{v}^{\dag }  
\\
\label{Hrad}
&\hat{H}_{rad} = \sum _{\alpha \in i,\{ f\} } \nu _{\alpha } \hat{a}_{\alpha }^{\dag } \hat{a}_{\alpha } ;          
\\ 
\label{Vrad} 
&\hat{V}_{rad} = \hat{V}_{rad}^{C} +\hat{V}_{rad}^{M}
\\
&= \sum _{J=C,M}\sum _{\alpha \in i,\{ f\} } \left(U_{gx,\alpha }^{J} (\hat{Q}_{v} ){\kern 1pt} \hat{D}_{J}^{\dag } \hat{a}_{\alpha } +U_{\alpha ,gx}^{J} (\hat{Q}_{v} ){\kern 1pt} \hat{a}_{\alpha }^{\dag } {\kern 1pt} \hat{D}_{J} \right)
\nonumber  
\end{align} 
$\hat{H}_{M} $ is the molecular Hamiltonian, modeled as a two electronic levels (ground and excited 
states, $g$ and $x$) system with a single molecular vibration of 
frequency $\omega _{v} $. $\hat{V}_{e-v} $ is the corresponding polaronic-type e-v coupling. 
$\hat{H}_{NS} $ describes a free electron metal and $\hat{V}_{NS} $ is the molecule-metal 
electron transfer coupling. $\hat{d}_{m}^{\dag } $ ($\hat{d}_{m} $) (\textit{m=g},
\textit{x}) and $\hat{c}_{k}^{\dag } $ ($\hat{c}_{k} $) create (annihilate) electrons in the molecular orbitals 
and the metal, respectively. $\hat{v}^{\dag } $ ($\hat{v}$) and$\hat{a}_{\alpha } ^{\dag }$ 
($\hat{a}_{\alpha } $) create (annihilate) vibrational excitations and photons in mode \textit{$\alpha $}, 
respectively. Finally, $\hat{D}_{M} =\hat{d}_{g}^{\dag } \hat{d}_{x} $ and 
$\hat{D}_{C} =\hat{d}_{g}^{\dag } \hat{c}_{k} $ are respectively intramolecular and charge-transfer 
electronic de-excitation operators.  $\hat{V}_{rad} $, 
Eq. \eqref{Vrad}, is the system-radiation field coupling in which the coupling matrix 
elements $U_{gx,\alpha }^{J} $ connect the molecular ground state \textit{g} to excited states \textit{x} in 
the channels $J=C\, \, {\rm or}\, \, M$ that determines the nature of the excited states \textit{x} 
(a similar term, not shown, is associated with light scattering from the bare metal).
As usual we distinguish between the incoming (occupied) photon mode $i$ and the 
continuum of accepting (empty) modes $\{ f\} $. These coupling elements are assumed to 
be functions of the vibrational coordinate, described here by their Taylor expansion up to linear term 
$(J=M,C)$
\begin{equation} \label{ZEqnNum982018} 
U_{gx,\alpha }^{J} (\hat{Q}_{v} )\equiv U_{\alpha ,gx}^{J*} (\hat{Q}_{v} )\approx U_{gx,\alpha }^{J0} 
+U_{gx,\alpha }^{J1} \hat{Q}_{v} 
\end{equation} 

To account for the observed dependence of the inelastic spectrum 
on the incident light intensity, we assume that in addition to being scattered, the incident radiation 
field affects the molecule-metal coupling. This can arise from a field-induced renormalization
of the electron transfer coupling $\hat{V}_{NS} $, Eq. \eqref{VNS},\cite{GanichevPRL98,CuevasNatelsonNN10,SelzerJPCL13}
as well as contributions from molecule to metal charge transfer transitions.
In the calculations reported below this is incorporated by an assumed dependence of the self-energies 
$\Gamma _{m}$ ($m=g,x$) on the incident light intensity. 

Consider first model A (Fig.~\ref{fig2}a). Here we assume that Raman scattering is dominated
by the intramolecular optical transition and disregard the corresponding contribution of the
optical charge transfer coupling $V^C_{rad}$.
This model can show interference between the vibrational and 
electronic scattering channels, provided that their coupling, Eq~(\ref{Vev}), 
is strong enough. 

Following the procedure described in Ref.~\onlinecite{GalperinRatnerNitzanJCP09}
the (normal) Raman scattering flux 
\footnote{We limit consideration to unbiased or weakly biased junctions, 
where only normal Raman scattering, where the initial 
electronic state is the ground state of the molecule, can take place.}
has the following structure
\begin{align} 
\label{Jif} 
&J_{i\to f} = \int _{-\infty }^{+\infty } d(t'-t)\int _{-\infty }^{0} d(t_{1} -t)\int _{-\infty }^{0} d(t_{2} -t')
\nonumber \\&
\times e^{-i\nu _{f} (t'-t)} {\kern 1pt} e^{-i\nu _{i} (t_{1} -t_{2} )}  
\\ & \times \left\langle \hat{U}_{i} (t_{2})\hat{D}(t_{2}){\kern 1pt} \hat{U}_{f} (t')
\hat{D}^{\dag } (t'){\kern 1pt} \hat{U}_{f} (t)\hat{D}(t){\kern 1pt} \hat{U}_{i} (t_{1} )\hat{D}^{\dag } (t_{1})\right\rangle
\nonumber
\end{align} 
in which the operators are summed over all relevant contributions from
Eqs.~\eqref{Vrad} and \eqref{ZEqnNum982018} and where 
the indices \textit{i} and \textit{f} 
indicate that the corresponding matrix element should be taken with the incident or scattered 
radiation field modes, respectively. In departure from Ref. [18] we keep only the lowest order 
(up to 2${}^{nd}$) terms in $U^{1} $, as usually done for non-resonant Raman calculations. 
The calculation proceeds as follows (See the SI for details): 

(a) Applying Eq. \eqref{Jif} to model A, yields Eqs.~(S2) for the light scattering flux. Analysis based on energy conservation suggests that expressions (S2a), (S2f) and (S2k) contribute to Rayleigh and pure electronic Raman scattering, while expressions (S2g) - (S2j) contribute to vibrational Raman scattering (possibly dressed by e-h excitations).\footnote{In particular, Eq. (S2a) leads to results discussed in our previous 
publications\cite{MGANJPCL11,MGANPRB11} on electronic Raman scattering, 
while (S2g) - (S2j) are analogs of the terms considered in our treatment\cite{GalperinRatnerNitzanNL09,GalperinRatnerNitzanJCP09} of (vibrational) resonant Raman scattering. Except that here we focus on the off-resonant scattering process.} 
Expressions (S2b) - (S2e) formally represent interference between vibrational and 
electronic-Raman/Rayleigh scattering. At the level of the present calculation, fourth order in 
the coupling to radiation field, they vanish in the absence of 
the coupling\eqref{Vev}.

(b) The e-v interaction \eqref{Vev} is considered to the lowest order. 
This introduces an additional term 
\begin{equation} \label{10)} 
\int _{c} d\tau _{v} {\kern 1pt} M_{g} {\kern 1pt} \hat{Q}_{v} (\tau _{v} ){\kern 1pt} \hat{n}_{g} (\tau _{v} ) 
\end{equation} 
into the correlation functions (S2b)-(S2e), yielding non-vanishing corrections 
that represent interference between the electronic and vibrational Raman processes 
that can give rise to Fano lineshape in the Raman scattering as shown below.\footnote{A closer look 
at the details of these corrections reveals that corrections to (S2b)-(S2c) from
projections (e)-(h) of Fig. S5 and corrections to (S2d)-(S2e) from projections (a)-(d) of Fig. S5
represent renormalization of the electronic Raman signal due to e-v interaction. These terms
dominate the resulting Fano resonance that is seen in the calculated Raman scattering 
at low incident lightintensity (inset to Fig. 3). Corrections to (S2b)-(S2c) from projections (a)-(d) 
of Fig. S5 and
corrections to (S2d)-(S2e) from projections (e)-(h) of Fig. S5 lead to renormalization of the
vibrational Raman scattering and are included as well in the calculations reported here.} 
In evaluating these terms, the real time analog, $t_{v} $, of the contour variable $\tau _{v} $ can be 
placed in all possible ways between the times $t$, $t'$, $t_{1} $, and $t_{2} $ on the Keldysh contour 
(see Fig.~S5).

(c) The light scattering flux including the interference corrections is thus obtained in terms of correlation 
functions of a system defined by the quadratic Hamiltonian $H_{0} $. Thus Wick's theorem applies and 
electronic and vibrational degrees of freedom decouple yielding the final expressions  
in terms of projections to real time of the electronic and vibrational Green functions, 
$
G_{mm'} (\tau ,\tau ')\equiv -i\langle T_{c} {\kern 1pt} \hat{d}_{m} (\tau ){\kern 1pt} \hat{d}_{m'}^{\dag } (\tau ')\rangle  
$ 
and 
$ 
D(\tau ,\tau ')\equiv -i\langle T_{c} {\kern 1pt} \hat{Q}_{v} (\tau ){\kern 1pt} \hat{Q}_{v} (\tau ')\rangle  
$, 
where $T_{c} $ is the contour ordering operator. For the molecular vibration we use the quasiparticle 
approximation, Eqs. (S3) and (S4). In evaluating the electronic GFs 
we assume  $\varepsilon _{x} -\varepsilon _{g} {\rm \gg }\Gamma _{m}$,
where $\Gamma _{m} =2\pi \sum _{k} |V_{km} |^{2} \delta (E-\varepsilon _{k})$; $m=g,x$.
Consequently, interstate correlations induced by the metal are neglected,
$G_{mm'}(\tau ,\tau ')\approx\delta _{m,m'}G_{mm}(\tau,\tau')\equiv\delta _{m,m'}G_{m}(\tau ,\tau ')$, leading to the standard expressions (S5)-(S7) for the electronic GFs. 

This procedure leads to explicit expressions, Eqs.~(S8)-(S14), for the different contributions 
to the light scattering flux in the relevant order (fourth) in the system-radiation field coupling
including the interference corrections, Eqs.~(S10) and (S14), arising from the e-v coupling. 
The other contributions can be classified according to their physical origin that 
can be identified by their energy conservation structure. 
Eq.~(S8d) and (S10d) represent the the Rayleigh scattering component that is disregarded 
in our calculation.
Eq. (S8) represents the pure electronic Raman that may be though of as an electronic 
sideband of the Rayleigh peak.\footnote{Eq.~(S8) is the pure electronic Raman induced by 
the optical charge transfer molecule-metal coupling. A similar term with similar spectral characteristics
corresponds to Raman scattering from the bare metal if the latter is structured enough 
(as opposed to a flat surface)} 
We have previously\cite{MGANJPCL11,MGANPRB11} used 
the Stokes (S) and anti-Stokes (AS) components 
of this contribution to estimate the apparent electronic heating in non-equilibrium molecular junctions. 
Eq. (S13) is the vibrational contribution to the Raman signal together with its own
electronic sideband.

\begin{figure}[t]
\centering\includegraphics[width=\linewidth]{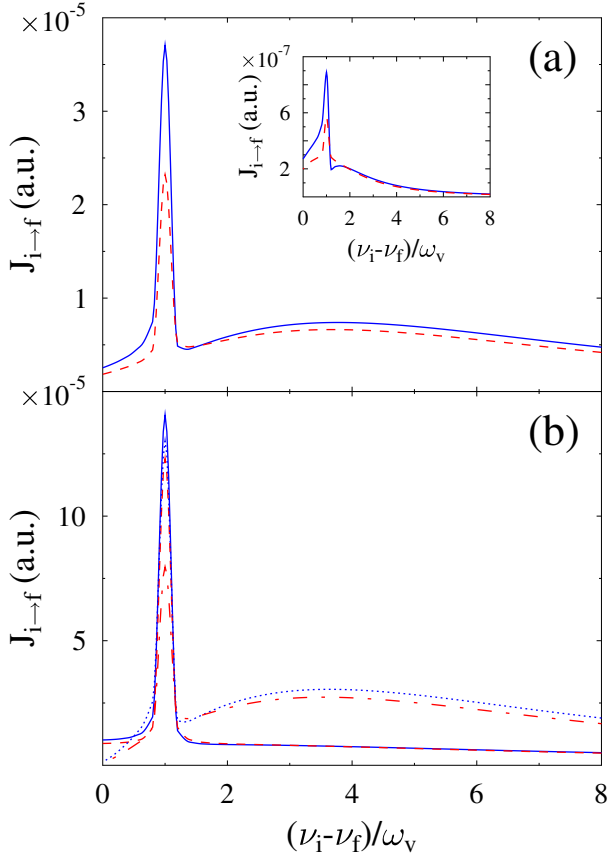}
\caption{\label{fig3} (Color online) 
(a) The total Raman scattering (full line; blue) evaluated from Eqs.~(S8)-(S14) 
(without the Rayleigh contributions, Eqs.~(S8d) and (S10d)), displayed against the Raman shift 
$\nu_{i}-\nu_{f}$.  The parameters used are $T=300$~K, $\varepsilon_{g} =0$, 
$\varepsilon_{x} =2$~eV, $\Gamma_{g}=\Gamma_{x}=0.05+\Gamma^{opt}$~eV with 
$\Gamma^{opt}=0.3$~eV being the light induced tunneling rate, $\omega_{v} =0.05$~eV, 
$M_{g}=0.02$~eV, $\nu _{i} =1$~eV, $U^{(0)}=0.1$~eV, $U^{(1)}=0.01$~eV. 
The red dashed line is the result obtained when the e-v interaction \eqref{Vev} is 
disregarded. The inset shows the corresponding weak incident field case with 
$\Gamma^{opt}=0.03$~eV, $U^{(0)}=0.032$~eV, $U^{(1)}=0.0032$~eV. 
(b) The Raman scattering calculated with the same parameters as in the main panel on the left 
at temperatures $3000$~K (solid and dashed lines) and $30$~K (dotted and dash-dotted lines, 
multiplied by a scale-factor $4$).
}
\end{figure}

Fig.~\ref{fig3} shows the total Raman scattering signal resulting from this model calculation. 
The calculation was done on an energy grid spanning the region from $-3$ to $3$ eV with step 
$10^{-3} $ eV and the molecular vibration is assumed to equilibrate quickly to the ambient temperature. 
The following observations can be made:

(a) Interference between electronic and vibrational inelastic scattering routes can indeed
lead to a Fano-type scattering spectrum. The electron-vibration coupling 
\eqref{Vev} is crucial for obtaining this behavior. 
This is seen in the weak illumination case (inset to Fig.~\ref{fig3}a) 
by comparing the results obtained with and without this coupling.

(b) The asymmetric lineshape that seen in the high incident intensity case (main figure~\ref{fig3}a) 
is not of this origin. In this case the calculation yields Fano-like lineshapes that qualitatively reproduce 
the experimental behavior also when the coupling \eqref{Vev} is disregarded. 

(c) The transition from a standard, essentially symmetric Raman 
line to a broadened assymetric line when the incident light intensity increases is best rationalized 
within this model by assuming that the coupling \eqref{Vev} is small enough
to be disregarded.  
This yields the dashed lines in Figure~\ref{fig3} that qualitatively agree with the experimental observation.

Our analysis thus shows that interference between vibrational and 
electronic Raman scattering can in principle occur leading to a characteristic Fano lineshape 
in Raman scattering from molecules adsorbed on metal substrates.
It suggests, however, that the origin of of the present observation of asymmetric Raman
lineshape at higher (photoinduced) metal-molecule coupling is different
resulting from the S component of the electronic Raman scattering. 
This is seen in the Fig.~\ref{fig3}b, that show that the asymmetry disappears 
at high ambient temperatures.

\begin{figure}[t]
\centering\includegraphics[width=\linewidth]{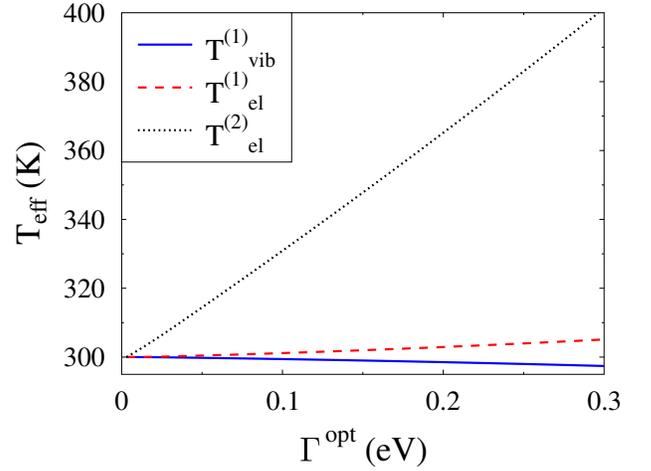}
\caption{\label{fig4} 
The effective temperature obtained from the Raman scattering spectrum displayed in 
the main panel of Fig. 3. See text for details.
}
\end{figure}

Our calculation also yields the effective (Raman) temperature from the ratio between the S 
and AS scattering intensities according to
\cite{NatelsonNatNano11,GalperinRatnerNitzanNL09,GalperinRatnerNitzanJCP09} 
$T_{eff}^{(1)}=\frac{\Delta \nu }{\ln \left(J_{\nu _{i} \to \nu _{i} -\Delta \nu } /J_{\nu _{i} \to \nu _{i} 
+\Delta _{\nu } } \right)}$, or fitting the tail of the AS signal to 
$a\,\Delta\nu/\big(1-\exp[-\Delta\nu/T_{eff}^{(2)}]\big)$, Eq.(S1),
where $\Delta \nu \equiv |\nu _{i} -\nu _{f} |$ is the Raman shift. 
For $T_{eff}^{(1)}$ we have arbitrarily assigned the vibrational Raman
temperature to the value calculated at $\Delta \nu =\omega _{v} $ and the electronic Raman 
temperature to that obtained at $\Delta \nu =2\omega _{v} $. 
The results for the parameters of Fig.~\ref{fig3}a (main panel) are shown in Fig.~\ref{fig4}. 
The resulting trends are similar to those observed experimentally, although the model calculation 
somewhat underestimates the electronic heating. 
In spite of this qualitative agreement with the experimental observation, 
we note that while $T_{eff}$ can give a rough estimate of system heating, 
the actual numbers may be meaningless. 
This is indicated here by discrepancy between the different calculations of $T_{el}$ 
and the apparent cooling seen in $T_{vib}$, keeping in mind that the oscillator was kept 
at the ambient temperature in this calculation.

\begin{figure}[t]
\centering\includegraphics[width=\linewidth]{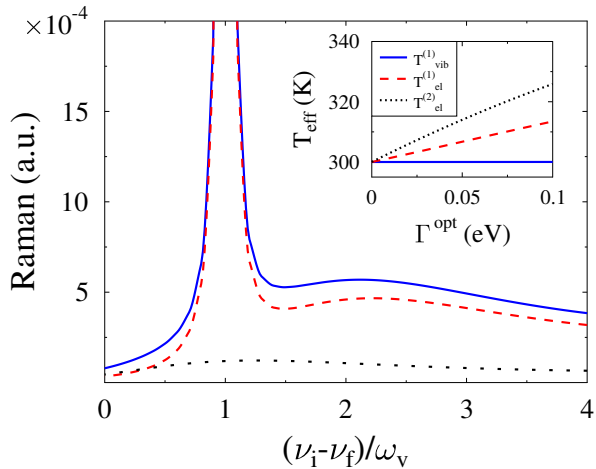}
\caption{\label{fig5} 
The Raman scattering signal calculated for model B (see text for parameters). 
The dotted-black and dashed-red lines represent the signals associated with 
the pure-electronic and electronic-dressed vibrational Raman, respectively, 
and the full blue line is their sum. 
The inset shows the corresponding vibrational and electronic effective temperatures, 
calculated as before.
}
\end{figure}

We have also considered model B as an alternative mechanism for interference.
The calculation follows the same steps as 
in (a) and (c) above, now using both M and C terms in Eq. \eqref{Vrad} and again 
keeping only terms up to the lowest order in $U^{1}$ (Eq. \eqref{ZEqnNum982018}). 
In this calculation we have restricted our considerations to contributions from the purely 
intra-molecular processes ($\hat{D}=\hat{D}_{M} $ in all four terms in \eqref{Jif}), 
purely charge-transfer processes ($\hat{D}=\hat{D}_{C}$ in all terms), 
and their mixtures (potential source of interference) in which excitation operators at times 
$t_{2} $ and $t'$ are of one kind, while those at times $t$ and $t_{1} $ of the other. 
Again the molecule-metal tunneling coupling $\Gamma _{m} $ is assumed to contain an incident field 
dependent contribution.\footnote{In the calculation displayed in Fig 4 we have assumed 
that $\Gamma^{opt} $ is dominated by the photo-induced charge transfer process.} 
The final result for the light scattering flux for this model is given by Eqs. (S15)-(S19). 

Fig.~\ref{fig5} shows the Stokes side of the Raman response obtained from this model 
using the following parameters: $T= 300$~K, $\varepsilon _{g} =0$~eV, 
$\varepsilon _{x} =2$~eV, $\omega _{v} =0.1$~eV, $\nu _{i} =1$~eV,
$\Gamma _{g} =\Gamma _{x} =0.15$~eV${}+\Gamma^{opt}$ 
with $\Gamma^{opt} =0.1$~eV, $U_{M\alpha }^{(0)}=0.1$~eV and 
$U_{M\alpha }^{(1)} =0.01$~eV ($\alpha=i,f$), and $E_{F} =0$. 
Calculations are performed on an energy grid spanning the region from $-5$ to $5$~eV with step 
$10^{-3}$~eV. 
Asymmetric Raman profiles are seen as before, but again we find (see SI) 
that while there is some contribution of interference between channels M and C, the main source of 
lineshape asymmetry originates again from the electronic sideband that dresses the 
Rayleigh and vibrational scattering peaks mainly on their blue side. 
The ``electronic heating'' obtained here is stronger than in model A, 
closer to the magnitude observed experimentally, 
but the actual numbers should again be considered cautiously.

 In conclusion, we have observed asymmetric lineshape features in Raman scattering from 
 bipyridyl ethylene molecules adsorbed on gold nanostructures, with the following characteristics: 
(a) Asymmetry increases with incident light intensity. 
(b) The vibrational temperature appears not 
to increase even at the highest intensity used, while the apparent electronic temperature increases 
by up to $\sim 600$ K. The Fano-like appearance of these lineshapes 
suggests the possibly implication of interference between different 
scattering pathways. Model calculations show that such interference,
leading to the observed asymmetry, is indeed possible in systems 
with strong enough electron-vibrational coupling. 
Our calculations suggest, however, that the observed lineshape 
asymmetry is dominated by electronic scattering sidebands that dress 
the Rayleigh and vibrational scattering peaks, 
and its  dependence on incident light intensity can be explained by an optically induced component
in the molecule-metal electron-transfer coupling. 
This model also yields vibrational and electronic ``Raman temperatures'' that are consistent 
with experimental observation, but may reflect the complex nature of the 
non-equlibrium response rather than the actual temperature.
Fano-type interference in Raman scattering has been shown to be a theoretical possibility.

\begin{acknowledgments} 
The Research of AN is supported by the Israel Science Foundation and by 
the US-Israel Binational Science Foundation. AN thanks the Theoretical Physics Group at 
the Free University of Berlin for hospitality.
MG gratefully acknowledges support by the DOE (Early Career Award, DE-SC0006422). 
The experimental work was carried out at the NSF Center for 
Chemistry at the Space-Time Limit (CHE-0802913). EH is supported by the
Academy of Finland (Decision No. 265502).
\end{acknowledgments}


\renewcommand{\theequation}{S\arabic{equation}}
\renewcommand{\thefigure}{S\arabic{figure}}
\setcounter{equation}{0}
\setcounter{figure}{0}

\begin{widetext}

\section{Supplementary Information}

\subsection{Experimental Details}
s outlined in the main text, the measurements are carried out on single silica encapsulated gold 
dumbbells, the transmission electron micrograph (TEM) of which is shown in the inset to Fig.~\ref{fig1}. 
The nanosphere diameter is 95$\pm$5 nm and the intersphere spacing prior to irradiation is 
$\sim 1$ nm.  
As molecular reporter, bipyridyl ethylene (BPE) is adsorbed on the gold spheres prior to encapsulation. 
The dumbbells are dispersed on a silicon nitride membrane (20 nm thick) of the TEM grid by drop 
casting them in a dilute solution. After mapping out locations and geometries of the nanostructures 
using a scanning electron microscope (SEM), Raman scattering measurements are carried out under 
an optical microscope, in the backscattering geometry, using an NA = 0.625 objective. The excitation 
source is a continuous wave diode laser, operating at $\lambda=532$~nm, 
which is resonant with the anti-bonding quadrupolar plasmon on these structures.\cite{MarhabaJPCC09,ApkarianACSNano12}
On these dumbbells, the binding quadrupolar plasmon and the binding dipolar plasmon 
resonances occur near 560 nm and 780 nm, respectively.\cite{MarhabaJPCC09} 
The molecular vibrational Raman spectra 
invariably appear over a background continuum (Fig.~\ref{fig1}b), which can also be seen on bare dumbbells 
(Fig. 1a). The molecular lines broaden asymmetrically as a function of irradiation intensity, to eventually 
coalesce into the asymmetric Fano-like profile shown in Fig.~\ref{fig1}c and 1d. We see variations in details on 
different particles, along with hysteretic response during intensity cycling due to evolution in the 
structure of the intersphere junction, which is verified through TEM. A rich variety of phenomena can be 
identified during fusion of the plasmonic junction when the incident light intensity increases.\cite{ApkarianACSNano12}
Here, we focus on the development of the Fano-like profile that possibly indicates interference between 
the electronic Raman scattering continuum on gold and the discrete resonances of the molecule. The 
temperature of the molecular vibrations is inferred from the usual ratio of Stokes/Anti-Stokes scattering. 
The electronic temperature is obtained from the anti-Stokes branch of the continuum, which for Raman 
shifts ${\mathcal E}=\omega_i-\omega_s>kT$, can be seen in Fig.~\ref{fig2} to decay exponentially, 
$\exp(-{\mathcal E}/kT_e)$, as expected for e-h Raman scattering that terminates on 
thermally populated holes. The spectrum is explicitly fitted to the joint density of states, 
which appears in first term of Eq.~(\ref{eq1}).  The principal experimental observations 
are summarized as follows:

\begin{figure}[htbp]
\centering\includegraphics[width=0.8\linewidth]{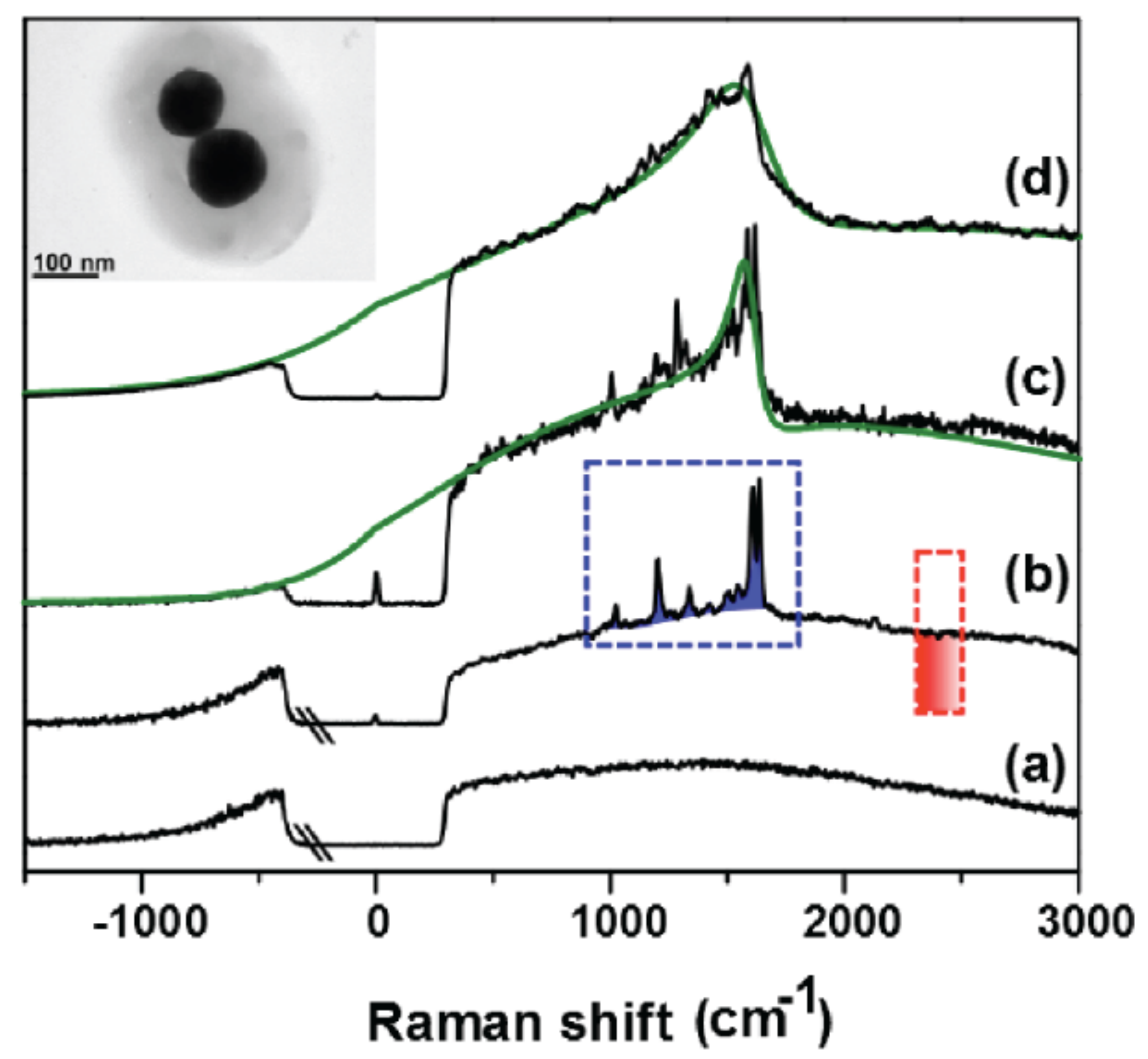}
\caption{\label{fig1}
(a) Raman spectrum recorded on a bare dumbbell, where reporter molecules are absent at 
the metallic junction. 
(b) Molecular Raman lines of BPE appear over the background continuum from a dumbbell. 
Both (a) and (b) have been recorded at low intensity ($\sim 10\mu$W/$\mu$m${}^{2}$) and 
anti-Stokes region is expanded vertically for clarity in either case (7x and 4.5x, respectively). 
At elevated irradiation intensity the entire spectrum develops Fano-like profile. 
Parentage of the molecular lines may still be identified (c) at moderate intensity 
($\sim 50 \mu$W/$\mu$m${}^{2}$) and becomes unidentifiable (d) at more intense irradiation 
($\sim150 \mu$W/$\mu$m${}^{2}$). The entire line profile in (c) and (d) can be fitted to 
Fano lineshape (green trace). The inset shows a TEM image of a typical dumbbell. 
Shaded regions are used to quantify the intensity dependence of molecular lines (blue) 
and electronic Raman scattering of gold (red), the results of which are shown in Fig.~\ref{fig1}b.
}
\end{figure}

\begin{figure}[htbp]
\centering\includegraphics[width=0.8\linewidth]{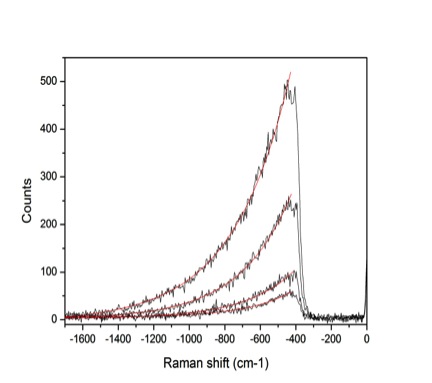}
\caption{\label{fig2}
The anti-Stokes branch of the continuum is given by the thermal occupation of the Fermi-Dirac distribution. This is illustrated by the exponential fits, 
$I_{AS}({\mathcal E})=\exp(-{\mathcal E}/kT_e)$, to the anti-Stokes continuum for four different irradiation intensities, $I=100, 50 , 25, 13 \mu$W/$\mu$m${}^2$, which yield electronic temperatures 
of $T_e = 465, 415, 395, 364$~K, respectively. A more accurate value of $T_e$  
is obtained from fitting to (\ref{eq1}), but the trend does not change. 
}
\end{figure}

\begin{enumerate}
\item  At low intensity, 10$\mu $W/$\mu $m${}^{2}$, the molecular vibrational Raman spectrum that 
appears over the background continuum in Fig.~\ref{fig1}b, perfectly matches that of the isolated BPE 
molecule. Note that the traces in Fig.~\ref{fig1} are identical to those in Fig.~1 of the main text. 
This has been shown in some detail 
previously.\cite{ApkarianNatPhoton14}
\item  The background continuum, which is also present on the bare dumbbell (Fig.~\ref{fig1}a), 
can be assigned to electronic Raman scattering (ERS) on the gold nano antenna. 
The continuum is polarized. As in the Hertzian antenna,\cite{EnghetaPRB08} the scattering is dipolar, 
with \textit{cos${}^{4}$q} polarization along the long axis of the dumbbell. 
\item  The fits of the anti-Stokes spectra to Eq.~\ref{eq1}), which are shown in Fig.~\ref{fig2}, 
yield the effective electronic temperature of the gold. 
This has been independently recognized recently.\cite{HugallBaumbergNL15}
Note, the method does not involve a ratio between Stokes and anti-Stokes scattering, 
which can be misleading. 
\textit{T${}_{e}$,} is a linear function of the irradiation intensity, as seen in Fig.~\ref{fig3}a.
\item  As the excitation intensity is increased up to 400 $\mu $W/$\mu $m${}^{2}$, the molecular lines 
develop asymmetric profiles and gain intensity relative to the background. At high intensity, the 
molecular response collapses into a single asymmetric line. The spectral profiles 
\ref{fig1}c and \ref{fig1}d can be fitted to a sum of ERS background and a Fano line, according to
\begin{equation}
\label{eq1}
W\left({\rm {\mathcal E}}\right)= a\int  f\left(E\right)\bigg(1-f\left(E+{\rm {\mathcal E}}\right)\bigg)e^{-\left|2{\rm {\mathcal E}}/{\rm d}\right|} dE
+ b\left[\frac{(q+\epsilon)^{2} }{1+\epsilon^{2} } -1\right]
\end{equation}
in which \textit{a,b} are normalization constants \textit{(a/b} =20), \textit{f(E)} is 
the Fermi-Dirac distribution, \textit{d} = 0.19 eV is the energy width of the projection of the collective plasmon state on single particle states,\cite{BanikPhD14} 
$\epsilon=\frac{\left({\mathcal E}-\omega_0\right)}{\gamma/2}$ 
is the reduced frequency with center on the molecular vibrational frequency, 
\textit{w${}_{0}$}${}_{ }$= 1600 cm${}^{-1}$. The resulting fits are shown as green traces overlapping 
the lines c and d in Fig.~\ref{fig1}. In Fig.~\ref{fig1}c the fit parameters are $\gamma=130$~cm${}^{-1}$ 
and \textit{q} = 2.5, 
while in Fig.~\ref{fig1}d  $\gamma=390$~cm${}^{-1}$ and \textit{q} = 2.7.
\item  While the ERS background shows linear dependence on irradiation intensity, the integrated area 
under the molecular lines is superlinear, as shown in Fig.~\ref{fig3}b.

\begin{figure}[htbp]
\centering\includegraphics[width=0.8\linewidth]{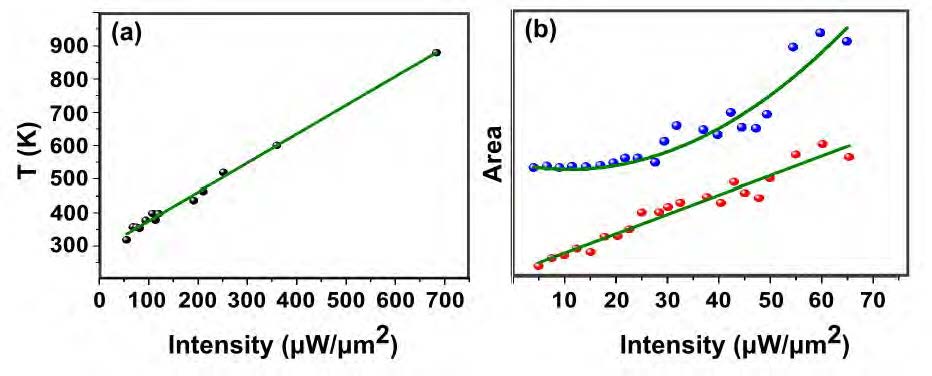}
\caption{\label{fig3}
(a) Exponentially decaying anti-Stokes branch of Raman spectra determines the temperature 
at the metallic junction which grows linearly with increasing irradiation intensity. 
(b) The electronic Raman scattering background (red dot) varies linearly with irradiation intensity. 
However, the area under the molecular lines (blue dot) is superlinear. 
}
\end{figure}

\item  Remarkably, the molecular vibrations and metal electrons appear to have different temperatures. 
At the highest incident light intensity, where the apparent electron temperature reaches 
$T_e=580$~K, we see no evidence of molecular anti-Stokes scattering. Based on 
the Stokes to anti-Stokes ratio, we establish that the apparent vibrational temperature of the molecule, 
$T_v$, is less than $300$~K.
\item  Near 1 mW/$\mu $m${}^{2}$, the nano-structure undergoes phase explosion -- the metal melts and the silica shell explodes. Hysteretic behavior is observed during intensity cycling at intensities in 
excess of $\sim 100 \mu$W/$\mu$m${}^{2}$, due to permanent evolution of the junction structure. 
The spectrum in Fig.~\ref{fig1}c, which was recorded at an irradiation intensity of 
$50 \mu$W/$\mu$m${}^{2}$, was obtained after such a cycle. 
\end{enumerate}

\subsection*{Theoretical Details}
Here we present details on derivation of Raman signal expressions for the two models considered 
in the paper. The starting point is general expression of the normal Raman scattering (see Eq.(9) 
of the paper and Fig.~\ref{fig4} below) \cite{GalperinRatnerNitzanJCP09,MGRatnerNitzanJCP15,MGRatnerNitzanARXIV15}.

Utilizing Taylor expansion of the optical electronic transition matrix element 
 (see Eq.(8) of the paper) and keeping terms up to second order 
in molecular vibration coordinate leads to
the following steady-state expression for Raman scattering 
from mode $i$ to mode $f$ of the radiation field
\begin{subequations}
\label{Jif}
\begin{align}
&J_{i\to f} = \int_{-\infty}^{+\infty} d(t'-t)\int_{-\infty}^{0} d(t_1-t)\int_{-\infty}^{0}d(t_2-t')\,
e^{i\nu_f(t-t')}-i\nu_i(t_1-t_2)
\nonumber \\ \label{Jifa} &\bigg[
U_{i,gx}^{(0)}U_{gx,f}^{(0)}U_{f,gx}^{(0)}U_{gx,i}^{(0)}
\langle\hat D(t_2)\hat Q_v(t_2)\,\hat D^\dagger(t')\,\hat D(t)\,\hat D^\dagger(t_1)\rangle
\\ \label{Jifb} & +
U_{i,gx}^{(1)}U_{gx,f}^{(0)}U_{f,gx}^{(0)}U_{gx,i}^{(0)}
\langle\hat D(t_2)\hat Q_v(t_2)\,\hat D^\dagger(t')\,\hat D(t)\,\hat D^\dagger(t_1)\rangle
\\ \label{Jifc} & +
U_{i,gx}^{(0)}U_{gx,f}^{(1)}U_{f,gx}^{(0)}U_{gx,i}^{(0)}
\langle\hat D(t_2)\,\hat D^\dagger(t')\hat Q_v^\dagger(t')\,\hat D(t)\,\hat D^\dagger(t_1)\rangle
\\ \label{Jifd} & +
U_{i,gx}^{(0)}U_{gx,f}^{(0)}U_{f,gx}^{(1)}U_{gx,i}^{(0)}
\langle\hat D(t_2)\,\hat D^\dagger(t')\,\hat D(t)\hat Q_v(t)\,\hat D^\dagger(t_1)\rangle
\\ \label{Jife} & +
U_{i,gx}^{(0)}U_{gx,f}^{(0)}U_{f,gx}^{(0)}U_{gx,i}^{(1)}
\langle\hat D(t_2)\,\hat D^\dagger(t')\,\hat D(t)\,\hat D^\dagger(t_1)\hat Q_v^\dagger(t_1)\rangle
\\ \label{Jiff} & +
U_{i,gx}^{(1)}U_{gx,f}^{(1)}U_{f,gx}^{(0)}U_{gx,i}^{(0)}
\langle\hat D(t_2)\hat Q_v(t_2)\,\hat D^\dagger(t')\hat Q_v^\dagger(t')\,\hat D(t)\,\hat D^\dagger(t_1)\rangle
\\ \label{Jifg} & +
U_{i,gx}^{(1)}U_{gx,f}^{(0)}U_{f,gx}^{(1)}U_{gx,i}^{(0)}
\langle\hat D(t_2)\hat Q_v(t_2)\,\hat D^\dagger(t')\,\hat D(t)\hat Q_v(t)\,\hat D^\dagger(t_1)\rangle
\\ \label{Jifh} & +
U_{i,gx}^{(1)}U_{gx,f}^{(0)}U_{f,gx}^{(0)}U_{gx,i}^{(1)}
\langle\hat D(t_2)\hat Q_v(t_2)\,\hat D^\dagger(t')\,\hat D(t)\,\hat D^\dagger(t_1)\hat Q_v^\dagger(t_1)\rangle
\\ \label{Jifi} & +
U_{i,gx}^{(0)}U_{gx,f}^{(1)}U_{f,gx}^{(1)}U_{gx,i}^{(0)}
\langle\hat D(t_2)\,\hat D^\dagger(t')\hat Q_v^\dagger(t')\,\hat D(t)\hat Q_v(t)\,\hat D^\dagger(t_1)\rangle
\\ \label{Jifj} & +
U_{i,gx}^{(0)}U_{gx,f}^{(1)}U_{f,gx}^{(0)}U_{gx,i}^{(1)}
\langle\hat D(t_2)\,\hat D^\dagger(t')\hat Q_v^\dagger(t')\,\hat D(t)\,\hat D^\dagger(t_1)\hat Q_v^\dagger(t_1)\rangle
\\ \label{Jifk} & +
U_{i,gx}^{(0)}U_{gx,f}^{(0)}U_{f,gx}^{(1)}U_{gx,i}^{(1)}
\langle\hat D(t_2)\,\hat D^\dagger(t')\,\hat D(t)\hat Q_v(t)\,\hat D^\dagger(t_1)\hat Q_v^\dagger(t_1)\rangle
\bigg]
\end{align}
\end{subequations}

\begin{figure}[t]
\centering\includegraphics[width=0.8\linewidth]{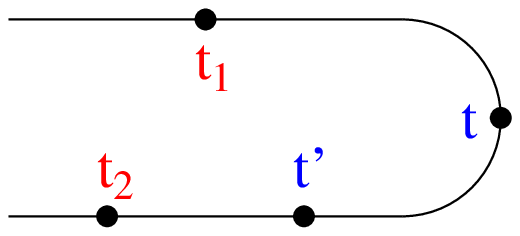}
\caption{\label{fig4}
(Color online) Projection for `the normal Raman' scattering. Times $t_1$ and $t_2$
indicate interaction with the incoming photon $\nu_i$,  $t$ and $t'$ - with the outgoing
photon $\nu_f$. The diagram is identical to Fig.~8a of Ref.~\cite{GalperinRatnerNitzanJCP09}.
}
\end{figure}

Analysis based on energy conservation suggests that from all the resulting diagrams
expressions (\ref{Jifa}), (\ref{Jiff}) and (\ref{Jifk}) contribute to electronic Raman scattering
(in particular, Eq.~(\ref{Jifa}) leads to results discussed in out previous 
publications\cite{MGANJPCL11,MGANPRB11}).
Expressions (\ref{Jifg})-(\ref{Jifj}) contribute to vibrational Raman scattering
(these results are similar to those of resonant Raman scattering consideration
in our previous publications\cite{GalperinRatnerNitzanNL09,GalperinRatnerNitzanJCP09}).
Finally, expressions (\ref{Jifb})-(\ref{Jife}) are responsible for interference between the
electronic and vibrational intra-molecular Raman scattering channels (see details below),
which is possible in the presence of electron-vibration interaction (see Eq.(5) of the paper).
This situation is encountered in description of the model A (see Fig.~2a of the paper).
If one disregards this interaction (and within the treatment of molecular coupling
to radiation field up to second order in molecular vibration coordinate) the latter contributions 
are zero. This is the situation encountered in description of the model B
(see Fig.~2b of the paper). We now turn to details of derivation specific for each of 
the models.

\subsubsection*{Model A}
Within this model we are interested in interference between electronic Raman scattering
(a process where vibrational state at the start an end of Raman scattering is the same)
and vibrational Raman scattering (a scattering process with initial and final states with
different vibrational excitations) amplitudes. For the two amplitudes to interfere,
final states of the two processes should be the same. Electron-vibration coupling 
in molecular Hamiltonian (see Eq.(5) of the paper), yields such a possibility (see Fig.~2a of the paper).
In the off-resonant Raman regime electron-vibration interaction can be disregarded. 
Utilizing the first order perturbation in the electron-vibration coupling 
(see Eq.(10) of the paper) in expressions (\ref{Jifb})-(\ref{Jife}) results in 
the desired interference. Possible placements of the time of electron-vibration  
interaction $t_v$ between times $t$, $t'$ and $t_1$,  $t_2$ representing interaction with 
the outgoing $f$ and incoming $i$ modes of the radiation field, respectively, 
are shown in  Fig.~\ref{fig5}. 

\begin{figure}[b]
\centering\includegraphics[width=\linewidth]{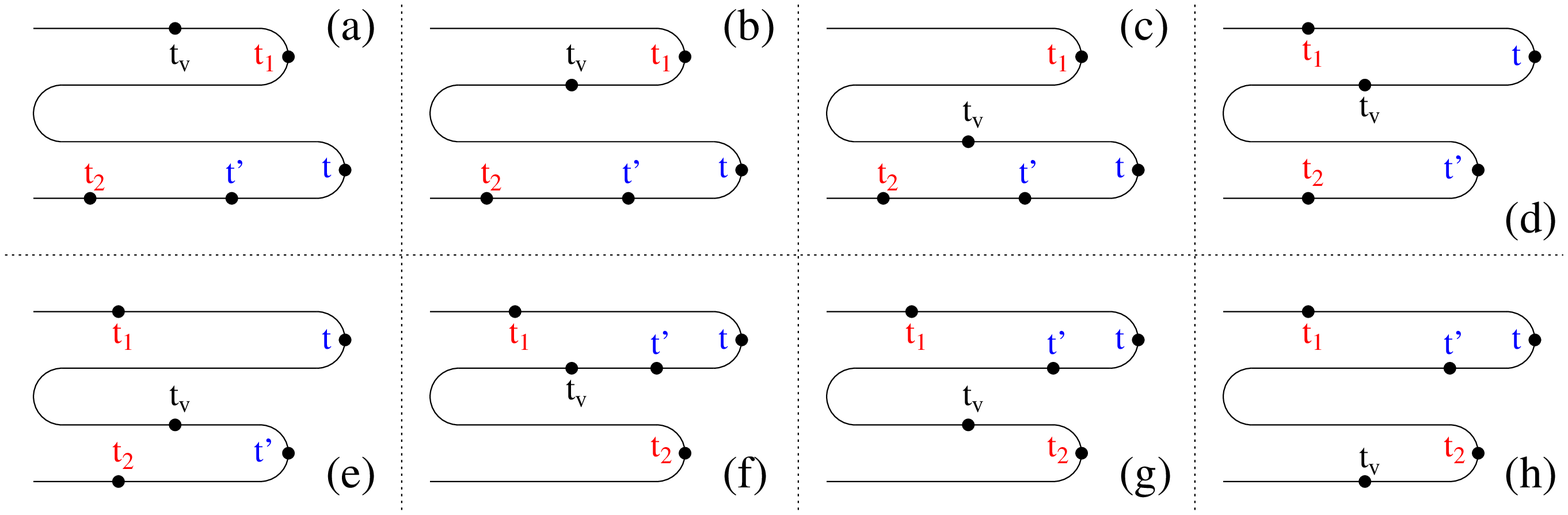}
\caption{\label{fig5}
(Color online) Electron-vibration interaction induced corrections to 
`the normal Raman' (Fig.~\ref{fig4}). $t_v$ indicates time of interaction with 
molecular vibration. The contour is deformed following the Langreth rules~\cite{HaugJauho_2008}.
}
\end{figure}

Projections (e)-(h) of Fig.~\ref{fig5} utilized in corrections to (\ref{Jifb})-(\ref{Jifc}) 
and projections (a)-(d) of Fig.~\ref{fig5} in corrections to (\ref{Jifd})-(\ref{Jife})
lead to renormalization of electronic Raman due to electron-vibration interaction.
These are the diagrams which lead to Fano resonance in Raman scattering (see below).
Projections (a)-(d) of Fig.~\ref{fig5} in corrections to (\ref{Jifb})-(\ref{Jifc}) 
and projections (e)-(h) of Fig.~\ref{fig5} in corrections to (\ref{Jifd})-(\ref{Jife})
renormalize vibrational Raman scattering. 

After accounting for the electron-vibration interaction we decouple electron and vibration
degrees of freedom and utilize the Wick's theorem to get final expressions in terms
of projections of electron and vibrational Green functions 
(see Eqs.~(11) and (12) of the paper).
We treat molecular vibration within quasiparticle approximation
\begin{align}
D^<(\omega) =& -2\pi i\left(N_v\delta(\omega-\omega_v)+[1+N_v]\delta(\omega+\omega_v)\right)
\\
D^{>}(\omega) =& -2\pi i\left(N_v\delta(\omega+\omega_v)+[1+N_v]\delta(\omega-\omega_v\right)
\end{align}
where $N_v$ is average population of the vibration.
Explicit expressions for electronic Green functions are ($m=g,x$)
\begin{align}
G^<_m(E) =& \frac{i\Gamma_m f(E)}{(E-\varepsilon_m)^2+(\Gamma_m/2)^2}
\\
G^>_m(E) =& \frac{-i\Gamma_m f(E)}{(E-\varepsilon_m)^2+(\Gamma_m/2)^2}
\\
G^r_m(E =& [E-\varepsilon_m+i\Gamma_m/2]^{-1}
\qquad G_m^a(E)=[G^r_m(E)]^{*}
\end{align}

After tedious but straightforward algebra we get the following steady-state expressions
for Raman scattering.
Explicit expression for the electronic Raman is
\begin{subequations}
\label{eR}
\begin{align}
\label{eRa}
& 
\int\frac{dE_{g1}}{2\pi}\int\frac{dE_{g2}}{2\pi}\int\frac{dE_{x1}}{2\pi}\int\frac{dE_{x2}}{2\pi}\,
2\pi\delta(\nu_{if}-E_{g21}-E_{x21}) 
\\ &
G_g^{<}(E_{g1})G_g^{>}(E_{g2})G_x^{<}(E_{x1})G_x^{>}(E_{x2})
\Bigg(\left\lvert\frac{U_{f,gx}^{(0)}U_{gx,i}^{(0)}}{\nu_i-E_{x2}+E_{g1}+i\delta}\right\rvert^2
\nonumber \\&
+2\mbox{Re}\, U_{i,gx}^{(1)}U_{gx,f}^{(1)}U_{f,gx}^{(0)}U_{gx,i}^{(0)}
\bigg[
\frac{1+N_v}{(\nu_i-E_{x2g1}-\omega_v+i\delta)(\nu_i-E_{x2g1}-i\delta)}
\nonumber \\ & \qquad\qquad\qquad\qquad\qquad\quad
+\frac{N_v}{(\nu_i-E_{x2g1}+\omega_v+i\delta)(\nu_i-E_{x2g1}-i\delta)}
\bigg]
\Bigg)
\nonumber \\ &
\label{eRb}
+\int\frac{dE_{x1}}{2\pi}\int\frac{dE_{x2}}{2\pi}
2\pi\delta(\nu_{if}-E_{x21})\, G_x^{<}(E_{x1})G_x^{>}(E_{x2})
\\ &
\Bigg(
 \left\lvert\int\frac{dE_{g}}{2\pi}
 \frac{U_{f,gx}^{(0)}U_{gx,i}^{(0)}G_g^{<}(E_g)}{\nu_i-E_{x2}+E_g+i\delta}\right\rvert^2
 -\int\frac{dE_{g1}}{2\pi}\int\frac{dE_{g2}}{2\pi}G_g^{<}(E_{g1})G_g^{<}(E_{g2})
\nonumber \\ & \times
2\mbox{Re}\, U_{i,gx}^{(1)}U_{gx,f}^{(1)}U_{f,gx}^{(0)}U_{gx,i}^{(0)}
\bigg[
\frac{1+N_v}{(\nu_i-E_{x2g1}-\omega_v+i\delta)(\nu_i-E_{x2g2}-i\delta)}
\nonumber \\ & \qquad\qquad\qquad\qquad\qquad\quad
+\frac{N_v}{(\nu_i-E_{x2g1}+\omega_v+i\delta)(\nu_i-E_{x2g2}-i\delta)}
\bigg]
\Bigg)
\nonumber \\ &
\label{eRc}
+\int\frac{dE_{g1}}{2\pi}\int\frac{dE_{g2}}{2\pi}
2\pi\delta(\nu_{if}-E_{g21})\, G_g^{<}(E_{g1})G_g^{>}(E_{g2})
\\ & 
\Bigg(
 \left\lvert\int\frac{dE_{x}}{2\pi}
 \frac{U_{f,gx}^{(0)}U_{gx,i}^{(0)}G_x^{>}(E_x)}{\nu_i-E_{x}+E_{g1}+i\delta}\right\rvert^2
 -\int\frac{dE_{x1}}{2\pi}\int\frac{dE_{x2}}{2\pi}G_x^{>}(E_{x1})G_x^{>}(E_{x2})
 \nonumber \\ & \times
2\mbox{Re}\, U_{i,gx}^{(1)}U_{gx,f}^{(1)}U_{f,gx}^{(0)}U_{gx,i}^{(0)}
\bigg[
\frac{1+N_v}{(\nu_i-E_{x1g1}-\omega_v+i\delta)(\nu_i-E_{x2g1}-i\delta)}
\nonumber \\ & \qquad\qquad\qquad\qquad\qquad\quad
+\frac{N_v}{(\nu_i-E_{x1g1}+\omega_v+i\delta)(\nu_i-E_{x2g1}-i\delta)}
\bigg]
\Bigg)
\nonumber \\ &
\label{eRd}
+ 2\pi\delta(\nu_i-\nu_f)\Bigg(
\left\lvert\int\frac{dE_{g}}{2\pi}\int\frac{dE_{x}}{2\pi}
\frac{U_{f,gx}^{(0)}U_{gx,i}^{(0)}G_g^{<}(E_g)G_x^{>}(E_x)}{\nu_i-E_x+E_g+i\delta}\right\rvert^2
\\ & \quad
+\int\frac{dE_{g1}}{2\pi}\int\frac{dE_{g2}}{2\pi}\int\frac{dE_{x1}}{2\pi}\int\frac{dE_{x2}}{2\pi}
G_g^{<}(E_{g1})G_g^{<}(E_{g2})G_x^{>}(E_{x1})G_x^{>}(E_{x2})
\nonumber \\ & \quad \times
2\mbox{Re}\, U_{i,gx}^{(1)}U_{gx,f}^{(1)}U_{f,gx}^{(0)}U_{gx,i}^{(0)}
\bigg[
\frac{1+N_v}{(\nu_i-E_{x1g1}-\omega_v+i\delta)(\nu_i-E_{x2g2}-i\delta)}
\nonumber \\ & \qquad\qquad\qquad\qquad\qquad\qquad
+\frac{N_v}{(\nu_i-E_{x1g1}+\omega_v+i\delta)(\nu_i-E_{x2g2}-i\delta)}
\bigg]
\Bigg)
\nonumber
\end{align}
\end{subequations}
Here
\begin{equation}
\nu_{if}\equiv\nu_i-\nu_f\quad E_{xm_xgm_g}\equiv E_{xm_x}-E_{gm_g}
\quad E_{m21}\equiv E_{m2}-E_{m1}\  (m=g,x)
\end{equation}

Corrections to the electronic Raman due to electron-vibration coupling are
\begin{subequations}
\begin{align}
& M_g\int\frac{dE_{g1}}{2\pi}\int\frac{dE_{g2}}{2\pi}\int\frac{dE_{x1}}{2\pi}\int\frac{dE_{x2}}{2\pi}
\nonumber \\ &
\label{ceRa}
\Bigg( 2\pi\delta(\nu_{if}-E_{g21}-E_{x21})\, 
G_g^{<}(E_{g1})G_g^{>}(E_{g2})G_x^{<}(E_{x1})G_x^{>}(E_{x2})
\\ & \times
2\mbox{Re}\bigg[ 
\binom{1+N_v}{N_v}
\frac{[U_{i,gx}^{(1)}U_{gx,f}^{(0)}G^r_g(E_{g2}\mp\omega_v)+U_{i,gx}^{(0)}U_{gx,f}^{(1)}G_g^a(E_{g1}\mp\omega_v)]U_{f,gx}^{(0)}U_{gx,i}^{(0)}}{(\nu_i-E_{x2g1}+i\delta)(\nu_i-E_{x2g1}\mp\omega_v-i\delta)}
\nonumber \\ &\qquad\qquad
+N_v\frac{[U_{i,gx}^{(0)}U_{gx,f}^{(1)}G^r_g(E_{g2}\mp\omega_v)+U_{i,gx}^{(1)}U_{gx,f}^{(0)}G_g^a(E_{g1}\mp\omega_v)]U_{f,gx}^{(0)}U_{gx,i}^{(0)}}{\lvert\nu_i-E_{x2g1}+i\delta\rvert^2}
\nonumber \\ &\qquad\qquad
-i\int\frac{dE_3}{2\pi} \frac{G_g^{<}(E_{g3})U_{f,gx}^{(0)}U_{gx,i}^{(0)}}{\nu_i-E_{x2g1}+i\delta}
\bigg(\frac{U_{i,gx}^{(1)}U_{gx,f}^{(0)}[F(E_{g2},E_{g3})-F(E_{g2},E_{g1})]}{\nu_i-E_{x2g1}-E_{g2g3}-i\delta}
\nonumber \\ & \qquad\qquad\qquad\qquad\qquad\qquad\qquad\qquad
+
\frac{U_{i,gx}^{(0)}U_{gx,f}^{(1)}[F(E_{g3},E_{g1})-F(E_{g2},E_{g1})]}{\nu_i-E_{x2g3}-i\delta}\bigg)
\nonumber \\ & \qquad\qquad
+\frac{[U_{i,gx}^{(1)}U_{gx,f}^{(0)}\Phi(E_{g1})+U_{i,gx}^{(0)}U_{gx,f}^{(1)}\Phi^{*}(E_{g2})]U_{f,gx}^{(0)}U_{gx,i}^{(0)}}{\lvert\nu_i-E_{x2g1}+i\delta\rvert^2}
\bigg]
\nonumber \\ &
\label{ceRb}
-2\pi\delta(\nu_{if}-E_{x21})\,
G_g^{<}(E_{g1})G_g^{<}(E_{g2})G_x^{<}(E_{x1})G_x^{>}(E_{x2})
\\ & \times
2\mbox{Re}\bigg[ 
\binom{1+N_v}{N_v}
\frac{[U_{i,gx}^{(1)}U_{gx,f}^{(0)}G^r_g(E_{g2}\mp\omega_v)+U_{i,gx}^{(0)}U_{gx,f}^{(1)}G_g^a(E_{g2}\mp\omega_v)]U_{f,gx}^{(0)}U_{gx,i}^{(0)}}{(\nu_i-E_{x2g1}+i\delta)(\nu_i-E_{x2g2}\mp\omega_v-i\delta)}
\nonumber \\ &\qquad\qquad
+N_v\frac{[U_{i,gx}^{(0)}U_{gx,f}^{(1)}G^r_g(E_{g2}\mp\omega_v)+U_{i,gx}^{(1)}U_{gx,f}^{(0)}G_g^a(E_{g2}\mp\omega_v)]U_{f,gx}^{(0)}U_{gx,i}^{(0)}}{(\nu_i-E_{x2g1}+i\delta)(\nu_i-E_{x2g2}-i\delta)}
\nonumber \\ &\qquad\qquad
+\frac{[U_{i,gx}^{(1)}U_{gx,f}^{(0)}\Phi(E_{g2})+U_{i,gx}^{(0)}U_{gx,f}^{(1)}\Phi^{*}(E_{g2})]U_{f,gx}^{(0)}U_{gx,i}^{(0)}}{(\nu_i-E_{x2g1}+i\delta)(\nu_i-E_{x2g2}-i\delta)}
\bigg]
\nonumber 
\end{align}
\begin{align} &
\label{ceRc}
-2\pi\delta(\nu_{if}-E_{g21})\, 
G_g^{<}(E_{g1})G_g^{>}(E_{g2})G_x^{>}(E_{x1})G_x^{>}(E_{x2})\,
2\mbox{Re}\bigg[ \binom{1+N_v}{N_v}
\\ & \times
\bigg(
\frac{U_{i,gx}^{(1)}U_{gx,f}^{(0)}G^r_g(E_{g2}\mp\omega_v)}{\nu_i-E_{x1g1}+i\delta}
+\frac{U_{i,gx}^{(0)}U_{gx,f}^{(1)}G_g^a(E_{g1}\mp\omega_v)}{\nu_i-E_{x1g2}+i\delta}
\bigg)\frac{U_{f,gx}^{(0)}U_{gx,i}^{(0)}}{\nu_i-E_{x2g1}\mp\omega_v-i\delta}
\nonumber \\ & \qquad
+N_v\frac{U_{f,gx}^{(0)}U_{gx,i}^{(0)}}{\nu_i-E_{x2g1}+i\delta}
\bigg(
\frac{U_{i,gx}^{(0)}U_{gx,f}^{(1)}G^r_g(E_{g2}\mp\omega_v)}{\nu_i-E_{x1g1}\mp\omega_v-i\delta}
+\frac{U_{i,gx}^{(1)}U_{gx,f}^{(0)}G_g^a(E_{g1}\mp\omega_v)}{\nu_i-E_{x1g1}-i\delta}
\bigg)
\nonumber \\ & \qquad
-i\int\frac{dE_3}{2\pi} \frac{G_g^{<}(E_{g3})U_{f,gx}^{(0)}U_{gx,i}^{(0)}}{\nu_i-E_{x1g1}+i\delta}
\bigg(\frac{U_{i,gx}^{(1)}U_{gx,f}^{(0)}[F(E_{g2},E_{g3})-F(E_{g2},E_{g1})]}{\nu_i-E_{x2g1}-E_{g2g3}-i\delta}
\nonumber \\ & \qquad\qquad\qquad\qquad\qquad\qquad\qquad
+
\frac{U_{i,gx}^{(0)}U_{gx,f}^{(1)}[F(E_{g3},E_{g1})-F(E_{g2},E_{g1})]}{\nu_i-E_{x2g3}-i\delta}\bigg)
\nonumber \\ & \qquad
+\frac{[U_{i,gx}^{(1)}U_{gx,f}^{(0)}\Phi(E_{g1})+U_{i,gx}^{(0)}U_{gx,f}^{(1)}\Phi^{*}(E_{g2})]U_{f,gx}^{(0)}U_{gx,i}^{(0)}}{(\nu_i-E_{x1g1}+i\delta)(\nu_i-E_{x2g1}-i\delta)}
\bigg]
\nonumber \\ &
\label{ceRd}
+2\pi\delta(\nu_i-\nu_f)\,
G_g^{<}(E_{g1})G_g^{<}(E_{g2})G_x^{>}(E_{x1})G_x^{>}(E_{x2})
\\ & \times
2\mbox{Re}\bigg[ 
\binom{1+N_v}{N_v}
\frac{[U_{i,gx}^{(1)}U_{gx,f}^{(0)}G^r_g(E_{g2}\mp\omega_v)+U_{i,gx}^{(0)}U_{gx,f}^{(1)}G_g^a(E_{g2}\mp\omega_v)]U_{f,gx}^{(0)}U_{gx,i}^{(0)}}{(\nu_i-E_{x1g1}+i\delta)(\nu_i-E_{x2g2}\mp\omega_v-i\delta)}
\nonumber \\ &\qquad\qquad
+N_v\frac{[U_{i,gx}^{(0)}U_{gx,f}^{(1)}G^r_g(E_{g2}\mp\omega_v)+U_{i,gx}^{(1)}U_{gx,f}^{(0)}G_g^a(E_{g2}\mp\omega_v)]U_{f,gx}^{(0)}U_{gx,i}^{(0)}}{(\nu_i-E_{x1g1}+i\delta)(\nu_i-E_{x2g2}-i\delta)}
\nonumber \\ &\qquad\qquad
+\frac{[U_{i,gx}^{(1)}U_{gx,f}^{(0)}\Phi(E_{g2})+U_{i,gx}^{(0)}U_{gx,f}^{(1)}\Phi^{*}(E_{g2})]U_{f,gx}^{(0)}U_{gx,i}^{(0)}}{(\nu_i-E_{x1g1}+i\delta)(\nu_i-E_{x2g2}-i\delta)}
\bigg]
\Bigg)
\nonumber
\end{align}
\end{subequations}
where upper (lower) row and sign correspond to Stokes (anti-Stokes) scattering channel,
repectively, and
\begin{align}
&F(E_1,E_2) \equiv \frac{1}{\omega_v-E_{21}+i\delta}+\frac{1}{\omega_v+E_{21}-i\delta} 
\\
&\Phi(E) \equiv i \int\frac{dE'}{2\pi}\bigg(
\frac{G_g^{>}(E')}{E-E'-\omega_v-i\delta}-\frac{G_g^{<}(E')}{E-E'+\omega_v-i\delta}
+\frac{2G_g^{<}(E')}{\omega_v}
\bigg)
\end{align}

At steady-state explicit expression for the vibrational Raman is
\begin{subequations}
\label{vR}
\begin{align}
\label{vRa}
& \int\frac{dE_{g1}}{2\pi}\int\frac{dE_{g2}}{2\pi}\int\frac{dE_{x1}}{2\pi}\int\frac{dE_{x2}}{2\pi}\,
\binom{1+N_v}{N_v}
\nonumber \\ &
\Bigg( 2\pi\delta(\nu_{if}-E_{g21}-E_{x21}\mp\omega_v)\,
G_g^{<}(E_{g1})G_g^{>}(E_{g2})G_x^{<}(E_{x1})G_x^{>}(E_{x2})
\\ &\times
\bigg[
2\mbox{Re}\frac{U_{i,gx}^{(0)}U_{gx,f}^{(1)}U_{f,gx}^{(0)}U_{gx,i}^{(1)}}{(\nu_i-E_{x2g1}+i\delta)(\nu_i-E_{x2g1}\mp\omega_v-i\delta)}
\nonumber \\ &\qquad
+ \frac{U_{i,gx}^{(1)}U_{gx,f}^{(0)}U_{f,gx}^{(0)}U_{gx,i}^{(1)}}{\lvert\nu_i-E_{x2g1}\mp\omega_v+i\delta\rvert^2}
+\frac{U_{i,gx}^{(0)}U_{gx,f}^{(1)}U_{f,gx}^{(1)}U_{gx,i}^{(0)}}{\lvert\nu_i-E_{x2g1}+i\delta\rvert^2}
\bigg]
\nonumber \\ & 
\label{vRb}
-2\pi\delta(\nu_{if}-E_{x21}\mp\omega_v)\,
G_g^{<}(E_{g1})G_g^{<}(E_{g2})G_x^{<}(E_{x1})G_x^{>}(E_{x2})
\\ &\times
\bigg[
2\mbox{Re}\frac{U_{i,gx}^{(0)}U_{gx,f}^{(1)}U_{f,gx}^{(0)}U_{gx,i}^{(1)}}{(\nu_i-E_{x2g2}+i\delta)(\nu_i-E_{x2g1}\mp\omega_v-i\delta)}
\nonumber \\ &\qquad
+ \frac{U_{i,gx}^{(1)}U_{gx,f}^{(0)}U_{f,gx}^{(0)}U_{gx,i}^{(1)}}{(\nu_i-E_{x2g2}\mp\omega_v+i\delta)(\nu_i-E_{x2g1}\mp\omega_v-i\delta)}
\nonumber \\ &\qquad
+\frac{U_{i,gx}^{(0)}U_{gx,f}^{(1)}U_{f,gx}^{(1)}U_{gx,i}^{(0)}}{(\nu_i-E_{x2g2}+i\delta)(\nu_i-E_{x2g1}-i\delta)}
\bigg]
\nonumber \\ &
\label{vRc}
-2\pi\delta(\nu_{if}-E_{g21}\mp\omega_v)\,
G_g^{<}(E_{g1})G_g^{>}(E_{g2})G_x^{>}(E_{x1})G_x^{>}(E_{x2})
\\ &\times
\bigg[
2\mbox{Re}\frac{U_{i,gx}^{(0)}U_{gx,f}^{(1)}U_{f,gx}^{(0)}U_{gx,i}^{(1)}}{(\nu_i-E_{x1g1}+i\delta)(\nu_i-E_{x2g1}\mp\omega_v-i\delta}
\nonumber \\ &\qquad
+ \frac{U_{i,gx}^{(1)}U_{gx,f}^{(0)}U_{f,gx}^{(0)}U_{gx,i}^{(1)}}{(\nu_i-E_{x1g1}\mp\omega_v+i\delta)(\nu_i-E_{x2g1}\mp\omega_v-i\delta)}
\nonumber \\ &\qquad
+\frac{U_{i,gx}^{(0)}U_{gx,f}^{(1)}U_{f,gx}^{(1)}U_{gx,i}^{(0)}}{(\nu_i-E_{x1g1}+i\delta)(\nu_i-E_{x2g1}-i\delta)}
\bigg]
\nonumber \\ &
\label{vRd}
+2\pi\delta(\nu_{if}\mp\omega_v)\,
G_g^{<}(E_{g1})G_g^{<}(E_{g2})G_x^{>}(E_{x1})G_x^{>}(E_{x2})
\\ &\times
\bigg[
2\mbox{Re}\frac{U_{i,gx}^{(0)}U_{gx,f}^{(1)}U_{f,gx}^{(0)}U_{gx,i}^{(1)}}{(\nu_i-E_{x1g2}+i\delta)(\nu_i-E_{x2g1}\mp\omega_v-i\delta)}
\nonumber \\ &\qquad
+ \frac{U_{i,gx}^{(1)}U_{gx,f}^{(0)}U_{f,gx}^{(0)}U_{gx,i}^{(1)}}{(\nu_i-E_{x1g2}\mp\omega_v+i\delta)(\nu_i-E_{x2g1}\mp\omega_v-i\delta)}
\nonumber \\ &\qquad
+\frac{U_{i,gx}^{(0)}U_{gx,f}^{(1)}U_{f,gx}^{(1)}U_{gx,i}^{(0)}}{(\nu_i-E_{x1g2}+i\delta)(\nu_i-E_{x2g1}-i\delta)}
\bigg]
\Bigg)
\nonumber
\end{align}
\end{subequations}

\begin{figure}[htbp]
\centering\includegraphics[width=0.8\linewidth]{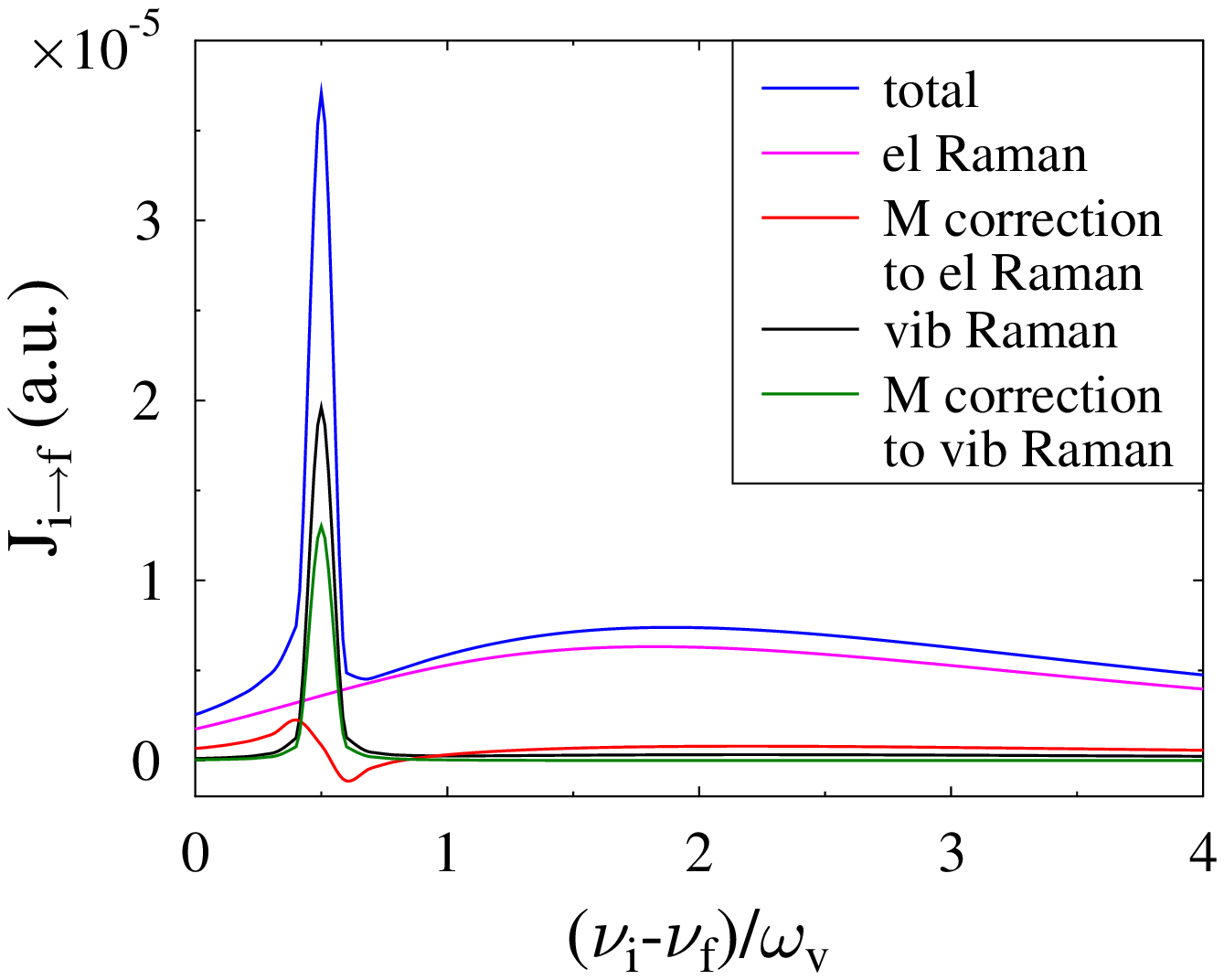}
\caption{\label{fig6}
(Color online) Intra-molecular Raman scattering (model A, see Fig.~3 of the paper).
Shown are the total signal (blue line), electronic Raman (magenta)  Eqs.~(\ref{eRa})-(\ref{eRd}) and its corrections due to electron-vibration interaction (red) 
Eqs.~(\ref{ceRa})-(\ref{ceRd}), as well as vibrational Raman (black)
Eqs.~(\ref{vRa})-(\ref{vRd}) and its corrections due to electron-vibration interaction (green) 
Eqs.~(\ref{cvRa})-(\ref{cvRd}).
}
\end{figure}

Corrections to the vibrational Raman due to electron-vibration coupling are
\begin{subequations}
\label{cvR}
\begin{align}
\label{cvRa}
& M_g \int\frac{dE_{g1}}{2\pi}\int\frac{dE_{g2}}{2\pi}\int\frac{dE_{x1}}{2\pi}\int\frac{dE_{x2}}{2\pi}\,
\binom{1+N_v}{N_v}
\nonumber \\ &
\Bigg( 2\pi\delta(\nu_{if}-E_{g21}-E_{x21}\mp\omega_v)\,
G_g^{<}(E_{g1})G_g^{>}(E_{g2})G_x^{<}(E_{x1})G_x^{>}(E_{x2})
\\ &\times
2\mbox{Re}\bigg[
\frac{G_g^r(E_{g1}\mp\omega_v)}{\nu_i-E_{x2g1}\mp\omega_v+i\delta}
\bigg(
\frac{U_{i,gx}^{(0)}U_{gx,f}^{(0)}U_{f,gx}^{(0)}U_{gx,i}^{(1)}}{\nu_i-E_{x2g1}\mp\omega_v-i\delta}
+\frac{U_{i,gx}^{(0)}U_{gx,f}^{(0)}U_{f,gx}^{(1)}U_{gx,i}^{(0)}}{\nu_i-E_{x2g1}-i\delta}
\bigg)
\nonumber \\ &\qquad
+\bigg(
\frac{U_{i,gx}^{(0)}U_{gx,f}^{(0)}U_{f,gx}^{(0)}U_{gx,i}^{(1)}}{\nu_i-E_{x2g1}\mp\omega_v-i\delta}
+\frac{U_{i,gx}^{(0)}U_{gx,f}^{(0)}U_{f,gx}^{(1)}U_{gx,i}^{(0)}}{\nu_i-E_{x2g1}-i\delta}
\bigg)
\frac{G_g^a(E_{g2}\pm\omega_v)}{\nu_i-E_{x2g1}+i\delta}
\bigg]
\nonumber \\ & 
\label{cvRb}
-2\pi\delta(\nu_{if}-E_{x21}\mp\omega_v)\,
G_g^{<}(E_{g1})G_g^{<}(E_{g2})G_x^{<}(E_{x1})G_x^{>}(E_{x2})
\\ &\times
2\mbox{Re}\bigg[
\frac{G_g^r(E_{g1}\mp\omega_v)}{\nu_i-E_{x2g1}\mp\omega_v+i\delta}
\bigg(
\frac{U_{i,gx}^{(0)}U_{gx,f}^{(0)}U_{f,gx}^{(0)}U_{gx,i}^{(1)}}{\nu_i-E_{x2g2}\mp\omega_v-i\delta}
+\frac{U_{i,gx}^{(0)}U_{gx,f}^{(0)}U_{f,gx}^{(1)}U_{gx,i}^{(0)}}{\nu_i-E_{x2g2}-i\delta}
\bigg)
\nonumber \\ &\qquad
+\bigg(
\frac{U_{i,gx}^{(0)}U_{gx,f}^{(0)}U_{f,gx}^{(0)}U_{gx,i}^{(1)}}{\nu_i-E_{x2g1}\mp\omega_v-i\delta}
+\frac{U_{i,gx}^{(0)}U_{gx,f}^{(0)}U_{f,gx}^{(1)}U_{gx,i}^{(0)}}{\nu_i-E_{x2g1}-i\delta}
\bigg)
\frac{G_g^a(E_{g2}\pm\omega_v)}{\nu_i-E_{x2g2}+i\delta}
\bigg]
\nonumber \\ & 
\label{cvRc}
-2\pi\delta(\nu_{if}-E_{g21}\mp\omega_v)\,
G_g^{<}(E_{g1})G_g^{>}(E_{g2})G_x^{>}(E_{x1})G_x^{>}(E_{x2})
\\ &\times
2\mbox{Re}\bigg[
\frac{G_g^r(E_{g1}\mp\omega_v)}{\nu_i-E_{x2g1}\mp\omega_v+i\delta}
\bigg(
\frac{U_{i,gx}^{(0)}U_{gx,f}^{(0)}U_{f,gx}^{(0)}U_{gx,i}^{(1)}}{\nu_i-E_{x1g1}\mp\omega_v-i\delta}
+\frac{U_{i,gx}^{(0)}U_{gx,f}^{(0)}U_{f,gx}^{(1)}U_{gx,i}^{(0)}}{\nu_i-E_{x1g1}-i\delta}
\bigg)
\nonumber \\ &\qquad
+\bigg(
\frac{U_{i,gx}^{(0)}U_{gx,f}^{(0)}U_{f,gx}^{(0)}U_{gx,i}^{(1)}}{\nu_i-E_{x2g1}\mp\omega_v-i\delta}
+\frac{U_{i,gx}^{(0)}U_{gx,f}^{(0)}U_{f,gx}^{(1)}U_{gx,i}^{(0)}}{\nu_i-E_{x2g1}-i\delta}
\bigg)
\frac{G_g^a(E_{g2}\pm\omega_v)}{\nu_i-E_{x1g1}+i\delta}
\bigg]
\nonumber \\ & 
\label{cvRd}
+2\pi\delta(\nu_{if}\mp\omega_v)\,
G_g^{<}(E_{g1})G_g^{<}(E_{g2})G_x^{>}(E_{x1})G_x^{>}(E_{x2})
\\ &\times
2\mbox{Re}\bigg[
\frac{G_g^r(E_{g1}\mp\omega_v)}{\nu_i-E_{x2g1}\mp\omega_v+i\delta}
\bigg(
\frac{U_{i,gx}^{(0)}U_{gx,f}^{(0)}U_{f,gx}^{(0)}U_{gx,i}^{(1)}}{\nu_i-E_{x1g2}\mp\omega_v-i\delta}
+\frac{U_{i,gx}^{(0)}U_{gx,f}^{(0)}U_{f,gx}^{(1)}U_{gx,i}^{(0)}}{\nu_i-E_{x1g2}-i\delta}
\bigg)
\nonumber \\ &\qquad
+\bigg(
\frac{U_{i,gx}^{(0)}U_{gx,f}^{(0)}U_{f,gx}^{(0)}U_{gx,i}^{(1)}}{\nu_i-E_{x2g1}\mp\omega_v-i\delta}
+\frac{U_{i,gx}^{(0)}U_{gx,f}^{(0)}U_{f,gx}^{(1)}U_{gx,i}^{(0)}}{\nu_i-E_{x2g1}-i\delta}
\bigg)
\frac{G_g^a(E_{g2}\pm\omega_v)}{\nu_i-E_{x1g2}+i\delta}
\bigg]
\Bigg)
\nonumber
\end{align}
\end{subequations}
Similar to our previous considerations~\cite{MGANJPCL11,MGANPRB11} 
we are interested in Raman scattering only.
Thus we drop Rayleigh contributions, Eqs.~(\ref{eRd}) and (\ref{ceRd}).
The latter are sharply peaked peaked at $\nu_i=\nu_f$.

Results of numerical simulations presented in Fig.~3 of the paper show asymmetric
scattering profile that becomes more pronounced at higher incident light intensity. Interference
between the electronic and vibrational scattering pathway contributes to this asymmetry as seen in
Fig.~\ref{fig6}. This interference is associated with the corrections to the electronic Raman due to
electron-vibration interaction, Eqs. (\ref{ceRa})-(\ref{ceRd}) 
(the orange line showing the correction due to
the coupling M to the electronic Raman scattering in Fig.~\ref{fig6}). 
However, the dominant contribution
to the line asymmetry is seen to be the Stokes nature of the electronic Raman scattering 
(purple line in Fig.~\ref{fig6}) that dresses the vibrational peak on its high energy side.


\subsubsection*{Model B}
Within the model B (see Fig.~2b of the paper) we focus on intra-molecular and charge transfer
contributions, and disregard electron-vibration coupling (see Eq.(5) of the paper). Thus
expressions (\ref{Jifb})-(\ref{Jife}), which are the source Fano-type interference in model A, 
do not contribute
in this model. Within the model radiation field leads to optical excitation of either intra-molecular
or charge transfer character. For each type of the excitations we separate vibrational and electronic
degrees of freedom in the correlation function, disregard coherence between ground and excited
states of the molecule, and perform integrals. A lengthy but straightforward derivations yield
explicit expressions for electronic and vibrational Raman signal. The steady-state Raman flux from
intial mode $i$ to final mode $f$ of the radiation field is

\begin{figure}[htbp]
\centering\includegraphics[width=0.8\linewidth]{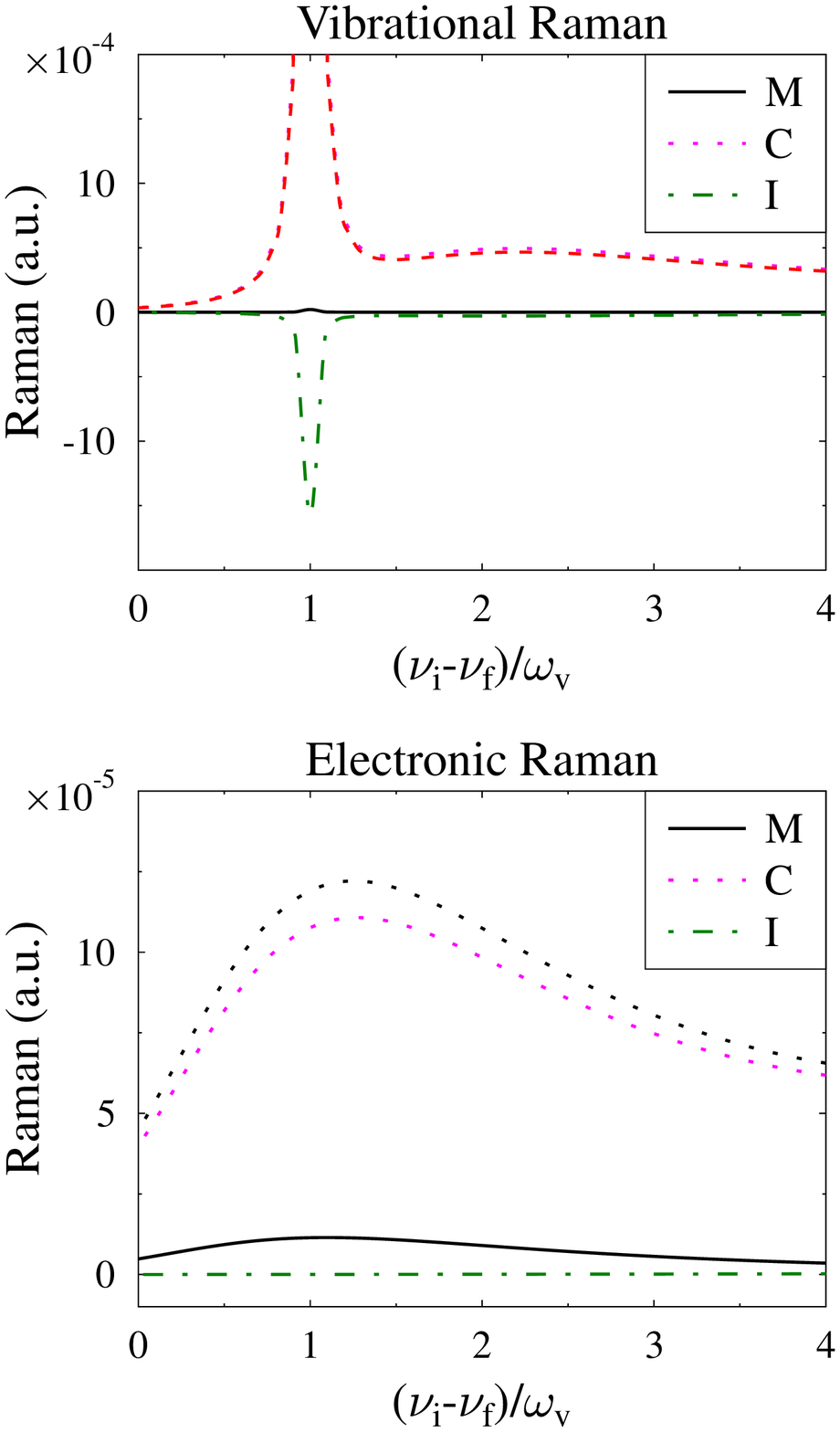}
\caption{\label{fig7}
(Color online) Vibrational (top) and electronic (bottom) Raman scattering 
(model B, see Fig.~5a of the paper).
Shown are the total signals (dashed red line - top graph, dotted black line - bottom graph),
as well as intra-molecular (solid black lines)  and charge-transfer (dotted magenta lines),
and their interferences (dash-dotted green lines).
}
\end{figure}

Steady-state Raman flux from initial mode $i$ to final mode $f$
of the radiation field is
\begin{equation}
 J_{i\to f} = \sum_{p=1}^4 J_{i\to f}^{(p)} 
\end{equation}
where
\begin{align}
\label{J1}
& J_{i\to f}^{(1)} = \sum_{s=\{0,-,+\}}
\int\frac{dE_g^{(1)}}{2\pi}\int\frac{dE_g^{(2)}}{2\pi}
\int\frac{dE_x^{(1)}}{2\pi}\int\frac{dE_x^{(2)}}{2\pi}
2\pi\delta(\nu_{if}-E_g^{(21)}-E_x^{(21)}+s\omega_v) N_s
\\
&\bigg[
 G_g^{<}(E_g^{(1)})G_g^{>}(E_g^{(2)})G_x^{<}(E_x^{(1)})G_x^{>}(E_x^{(2)})
\left\lvert
 \frac{U_{fM}^{(1)}U_{Mi}^{(0)}}{\nu_i-E_x^{(2)}+E_g^{(1)}+i\delta}
+\frac{U_{fM}^{(0)}U_{Mi}^{(1)}}{\nu_i-E_x^{(2)}+E_g^{(1)}+s\omega_v+i\delta}
\right\rvert^2
\nonumber \\
& +\sum_{k_1,k_2}G_g^{<}(E_g^{(1)})G_g^{>}(E_g^{(2)})g_{k_1}^{<}(E_x^{(1)})g_{k_2}^{>}(E_x^{(2)})
\left\lvert
 \frac{U_{fC}^{(1)}U_{Ci}^{(0)}}{\nu_i-E_x^{(2)}+E_g^{(1)}+i\delta}
+\frac{U_{fC}^{(0)}U_{Ci}^{(1)}}{\nu_i-E_x^{(2)}+E_g^{(1)}+s\omega_v+i\delta}
\right\rvert^2
\nonumber \\
& +2\mbox{Re}\sum_{k_1,k_2}G_g^{<}(E_g^{(1)})G_g^{>}(E_g^{(2)})G_{k_1x}^{<}(E_x^{(1)})G_{xk_2}^{>}(E_x^{(2)})
\nonumber \\
&\times\left(
 \frac{U_{iM}^{(1)}U_{Mf}^{(0)}U_{fC}^{(1)}U_{Ci}^{(0)}}
 {(\nu_i-E_x^{(2)}+E_g^{(1)}+i\delta)(\nu_i-E_x^{(2)}+E_g^{(1)}+s\omega_v-i\delta)}
+ \frac{U_{iM}^{(1)}U_{Mf}^{(0)}U_{fC}^{(0)}U_{Ci}^{(1)}}
 {\lvert\nu_i-E_x^{(2)}+E_g^{(1)}+s\omega_v+i\delta\rvert^2}\right.
\nonumber \\ &\left.\qquad
+ \frac{U_{iM}^{(0)}U_{Mf}^{(1)}U_{fC}^{(1)}U_{Ci}^{(0)}}
 {\lvert\nu_i-E_x^{(2)}+E_g^{(1)}+i\delta\rvert^2}
+\frac{U_{iM}^{(0)}U_{Mf}^{(1)}U_{fC}^{(0)}U_{Ci}^{(1)}}
 {(\nu_i-E_x^{(2)}+E_g^{(1)}+s\omega_v+i\delta)(\nu_i-E_x^{(2)}+E_g^{(1)}-i\delta)}
\right)
\bigg]
\nonumber 
\end{align}
\begin{align}
\label{J2}
& J_{i\to f}^{(2)} = -\sum_{s=\{0,-,+\}}
\int\frac{dE_g^{(1)}}{2\pi}\int\frac{dE_g^{(2)}}{2\pi}
\int\frac{dE_x^{(1)}}{2\pi}\int\frac{dE_x^{(2)}}{2\pi}\bigg(
2\pi\delta(\nu_{if}-E_x^{(21)}+s\omega_v) N_s
\\ &\bigg[
 G_g^{<}(E_g^{(1)})G_g^{<}(E_g^{(2)})G_x^{<}(E_x^{(1)})G_x^{>}(E_x^{(2)})
\left\lvert
 \frac{U_{fM}^{(1)}U_{Mi}^{(0)}}{\nu_i-E_x^{(2)}+E_g^{(1)}+i\delta}
+\frac{U_{fM}^{(0)}U_{Mi}^{(1)}}{\nu_i-E_x^{(2)}+E_g^{(2)}+s\omega_v+i\delta}
\right\rvert^2
\nonumber \\ &
+\sum_{k_1,k_2}G_g^{<}(E_g^{(1)})G_g^{<}(E_g^{(2)})g_{k_1}^{<}(E_x^{(1)})g_{k_2}^{>}(E_x^{(2)})
\left\lvert
 \frac{U_{fC}^{(1)}U_{Ci}^{(0)}}{\nu_i-E_x^{(2)}+E_g^{(1)}+i\delta}
+\frac{U_{fC}^{(0)}U_{Ci}^{(1)}}{\nu_i-E_x^{(2)}+E_g^{(2)}+s\omega_v+i\delta}
\right\rvert^2
\nonumber \\ &
+2\mbox{Re}\sum_{k_1,k_2}G_g^{<}(E_g^{(1)})G_g^{<}(E_g^{(2)})G_{k_1x}^{<}(E_x^{(1)})G_{xk_2}^{>}(E_x^{(2)})
\nonumber \\ &\times
\left(
 \frac{U_{iM}^{(1)}U_{Mf}^{(0)}U_{fC}^{(1)}U_{Ci}^{(0)}}
 {(\nu_i-E_x^{(2)}+E_g^{(1)}+i\delta)(\nu_i-E_x^{(2)}+E_g^{(2)}+s\omega_v-i\delta)}
\right. \nonumber \\ &\qquad
+ \frac{U_{iM}^{(1)}U_{Mf}^{(0)}U_{fC}^{(0)}U_{Ci}^{(1)}}
 {(\nu_i-E_x^{(2)}+E_g^{(1)}+s\omega_v+i\delta)(\nu_i-E_x^{(2)}+E_g^{(2)}+s\omega_v-i\delta)}
\nonumber \\ & \qquad
+ \frac{U_{iM}^{(0)}U_{Mf}^{(1)}U_{fC}^{(1)}U_{Ci}^{(0)}}
 {(\nu_i-E_x^{(2)}+E_g^{(1)}+i\delta)(\nu_i-E_x^{(2)}+E_g^{(2)}-i\delta)}
 \nonumber \\ &\qquad \left.
+\frac{U_{iM}^{(0)}U_{Mf}^{(1)}U_{fC}^{(0)}U_{Ci}^{(1)}}
 {(\nu_i-E_x^{(2)}+E_g^{(2)}+s\omega_v+i\delta)(\nu_i-E_x^{(2)}+E_g^{(1)}-i\delta)}
\right) \bigg]
\nonumber
\end{align}
\begin{align}
\label{J3}
& J_{i\to f}^{(3)} = -\sum_{s=\{0,-,+\}}
\int\frac{dE_g^{(1)}}{2\pi}\int\frac{dE_g^{(2)}}{2\pi}
\int\frac{dE_x^{(1)}}{2\pi}\int\frac{dE_x^{(2)}}{2\pi}\bigg(
2\pi\delta(\nu_{if}-E_g^{(21)}+s\omega_v) N_s
 \\ & \bigg[
 G_g^{<}(E_g^{(1)})G_g^{>}(E_g^{(2)})G_x^{>}(E_x^{(1)})G_x^{>}(E_x^{(2)})
\left\lvert
 \frac{U_{fM}^{(1)}U_{Mi}^{(0)}}{\nu_i-E_x^{(1)}+E_g^{(1)}+i\delta}
+\frac{U_{fM}^{(0)}U_{Mi}^{(1)}}{\nu_i-E_x^{(2)}+E_g^{(1)}+s\omega_v+i\delta}
\right\rvert^2
\nonumber \\ &
+\sum_{k_1,k_2}G_g^{<}(E_g^{(1)})G_g^{>}(E_g^{(2)})g_{k_1}^{>}(E_x^{(1)})g_{k_2}^{>}(E_x^{(2)})
\left\lvert
 \frac{U_{fC}^{(1)}U_{Ci}^{(0)}}{\nu_i-E_x^{(1)}+E_g^{(1)}+i\delta}
+\frac{U_{fC}^{(0)}U_{Ci}^{(1)}}{\nu_i-E_x^{(2)}+E_g^{(1)}+s\omega_v+i\delta}
\right\rvert^2
\nonumber \\ &
+2\mbox{Re}\sum_{k}G_g^{<}(E_g^{(1)})G_g^{>}(E_g^{(2)})g_{k}^{>}(E_x^{(1)})G_x^{>}(E_x^{(2)})
\nonumber \\ &\times
\left(
 \frac{U_{iM}^{(1)}U_{Mf}^{(0)}U_{fC}^{(1)}U_{Ci}^{(0)}}
 {(\nu_i-E_x^{(1)}+E_g^{(1)}+i\delta)(\nu_i-E_x^{(2)}+E_g^{(1)}+s\omega_v-i\delta)}
 \right. \nonumber \\ &\qquad
+ \frac{U_{iM}^{(1)}U_{Mf}^{(0)}U_{fC}^{(0)}U_{Ci}^{(1)}}
 {(\nu_i-E_x^{(1)}+E_g^{(1)}+s\omega_v+i\delta)(\nu_i-E_x^{(2)}+E_g^{(1)}+s\omega_v-i\delta)}
 \nonumber \\ &\qquad
+ \frac{U_{iM}^{(0)}U_{Mf}^{(1)}U_{fC}^{(1)}U_{Ci}^{(0)}}
 {(\nu_i-E_x^{(1)}+E_g^{(1)}+i\delta)(\nu_i-E_x^{(2)}+E_g^{(1)}-i\delta)}
 \nonumber \\ & \qquad \left.
+\frac{U_{iM}^{(0)}U_{Mf}^{(1)}U_{fC}^{(0)}U_{Ci}^{(1)}}
 {(\nu_i-E_x^{(1)}+E_g^{(1)}+s\omega_v+i\delta)(\nu_i-E_x^{(2)}+E_g^{(1)}-i\delta)}
\right)
\bigg]
\nonumber 
\end{align}
\begin{align}
\label{J4}
& J_{i\to f}^{(4)} = \sum_{s=\{-,+\}}
\int\frac{dE_g^{(1)}}{2\pi}\int\frac{dE_g^{(2)}}{2\pi}
\int\frac{dE_x^{(1)}}{2\pi}\int\frac{dE_x^{(2)}}{2\pi}\bigg(
2\pi\delta(\nu_{if}+s\omega_v) N_s
\\ &\bigg[
 G_g^{<}(E_g^{(1)})G_g^{<}(E_g^{(2)})G_x^{>}(E_x^{(1)})G_x^{>}(E_x^{(2)})
\left\lvert
 \frac{U_{fM}^{(1)}U_{Mi}^{(0)}}{\nu_i-E_x^{(1)}+E_g^{(2)}+i\delta}
+\frac{U_{fM}^{(0)}U_{Mi}^{(1)}}{\nu_i-E_x^{(2)}+E_g^{(1)}+s\omega_v+i\delta}
\right\rvert^2
\nonumber \\ &
+\sum_{k_1,k_2}G_g^{<}(E_g^{(1)})G_g^{>}(E_g^{(2)})g_{k_1}^{>}(E_x^{(1)})g_{k_2}^{>}(E_x^{(2)})
\left\lvert
 \frac{U_{fC}^{(1)}U_{Ci}^{(0)}}{\nu_i-E_x^{(1)}+E_g^{(2)}+i\delta}
+\frac{U_{fC}^{(0)}U_{Ci}^{(1)}}{\nu_i-E_x^{(2)}+E_g^{(1)}+s\omega_v+i\delta}
\right\rvert^2
\nonumber \\ &
+2\mbox{Re}\sum_{k}G_g^{<}(E_g^{(1)})G_g^{<}(E_g^{(2)})g_{k}^{>}(E_x^{(1)})G_x^{>}(E_x^{(2)})
\nonumber \\ &\times
\left(
 \frac{U_{iM}^{(1)}U_{Mf}^{(0)}U_{fC}^{(1)}U_{Ci}^{(0)}}
 {(\nu_i-E_x^{(1)}+E_g^{(2)}+i\delta)(\nu_i-E_x^{(2)}+E_g^{(1)}+s\omega_v-i\delta)}
 \right. \nonumber \\ &\qquad
+ \frac{U_{iM}^{(1)}U_{Mf}^{(0)}U_{fC}^{(0)}U_{Ci}^{(1)}}
 {(\nu_i-E_x^{(1)}+E_g^{(2)}+s\omega_v+i\delta)(\nu_i-E_x^{(2)}+E_g^{(1)}+s\omega_v-i\delta)}
 \nonumber \\ &\qquad
+ \frac{U_{iM}^{(0)}U_{Mf}^{(1)}U_{fC}^{(1)}U_{Ci}^{(0)}}
 {(\nu_i-E_x^{(1)}+E_g^{(2)}+i\delta)(\nu_i-E_x^{(2)}+E_g^{(1)}+i\delta)}
 \nonumber \\ & \qquad \left.
+\frac{U_{iM}^{(0)}U_{Mf}^{(1)}U_{fC}^{(0)}U_{Ci}^{(1)}}
 {(\nu_i-E_x^{(1)}+E_g^{(2)}+s\omega_v+i\delta)(\nu_i-E_x^{(2)}+E_g^{(1)}-i\delta)}
\right)
\bigg]
\nonumber
\end{align}
Here $\nu_{if}\equiv\nu_i-\nu_f$, $E_g^{(21)}\equiv E_g^{(2)}-E_g^{(1)}$,
$E_x^{(21)}\equiv E_x^{(2)}-E_x^{(1)}$, $G_{g(x)}^{>(<)}$ is greater (lesser)
projection of the molecular Green function, 
$g_k^{>(<)}$ is greater (lesser) projection of Green function
representing free electrons in nanoparticles
\begin{equation}
 g_k(\tau,\tau')\equiv -i\langle T_c\,\hat c_k(\tau)\,\hat c_k^\dagger(\tau')\rangle,
\end{equation}
and $G_{kx}^{>(<)}$ and $G_{xk}^{>(<)}$ are greater (lesser) projections 
of the mixed (molecule-nanoparticle) Green functions
\begin{align}
 G_{kx}(\tau,\tau')\equiv& -i\langle T_c\,\hat c_k(\tau)\,\hat d_x^\dagger(\tau')\rangle
 \\
 G_{xk}(\tau,\tau')\equiv& -i\langle T_c\,\hat d_x(\tau)\,\hat c_k^\dagger(\tau')\rangle
\end{align}
In Eqs.~(\ref{J1})-(\ref{J4}) sum over $s$ (${}=\{0,-,+\}$) corresponds to 
electronic ($s=0$) and Stokes ($s=-$) and anti-Stokes ($s=+$) 
vibrational contributions to the total Raman signal with
\begin{equation}
\label{Ns}
 N_s=\begin{cases}
 1 & s=0 \\
 N_v+1 & s=- \\
 N_v & s=+
 \end{cases}
\end{equation}
($N_v$ is average population of the vibrational mode).
Note that Eq.~(\ref{J4}) contributes to vibrational Raman only. 
Its electronic part contributes 
to Rayleigh scattering~\cite{MGANJPCL11,MGANPRB11},
and is disregarded in our consideration.
Each of four expressions (S16)-(S19) contains 
the pure contribution of channel M in their first row (these terms represent the equivalent of model A, 
but without the effect of the e-v coupling (5) and the pure contribution of 
channel C in their second row. The rest of these expressions correspond to interference between 
these channels. It should be emphasized that each of these contributions has in turn a vibrational 
Raman and electronic Raman components, characterized by the corresponding change 
(vibrational or electronic) in system state. 

Numerical simulations presented in Fig.~5 of the paper show that this model is also capable
of providing an asymmetric feature in the Stokes line. A closer examination of intra-molecular and
charge-transfer contributions as well as their interference to vibrational and electronic Raman
scattering (see Fig.~\ref{fig7}) indicate that the Fano-like feature in the total signal is mostly due 
to charge transfer
contribution to the vibrational Raman scattering. This contribution consists of multiple
scattering paths from different electronic states in the contact and in this sense it is an interference
feature. However, as in model A, the scattering lineshape asymmetry is dominated by the
electronic Stokes sideband dressing the molecular vibrational line.

\end{widetext}


\begin{thebibliography}{42}
\expandafter\ifx\csname natexlab\endcsname\relax\def\natexlab#1{#1}\fi
\expandafter\ifx\csname bibnamefont\endcsname\relax
  \def\bibnamefont#1{#1}\fi
\expandafter\ifx\csname bibfnamefont\endcsname\relax
  \def\bibfnamefont#1{#1}\fi
\expandafter\ifx\csname citenamefont\endcsname\relax
  \def\citenamefont#1{#1}\fi
\expandafter\ifx\csname url\endcsname\relax
  \def\url#1{\texttt{#1}}\fi
\expandafter\ifx\csname urlprefix\endcsname\relax\def\urlprefix{URL }\fi
\providecommand{\bibinfo}[2]{#2}
\providecommand{\eprint}[2][]{\url{#2}}

\bibitem[{\citenamefont{Moskovits}(1985)}]{MoskovitsRMP85}
\bibinfo{author}{\bibfnamefont{M.}~\bibnamefont{Moskovits}},
  \bibinfo{journal}{Rev. Mod. Phys.} \textbf{\bibinfo{volume}{57}},
  \bibinfo{pages}{783} (\bibinfo{year}{1985}).

\bibitem[{\citenamefont{Gersten and Nitzan}(1985)}]{GerstenNitzanSurfSci85}
\bibinfo{author}{\bibfnamefont{J.~I.} \bibnamefont{Gersten}} \bibnamefont{and}
  \bibinfo{author}{\bibfnamefont{A.}~\bibnamefont{Nitzan}},
  \bibinfo{journal}{Surface Science} \textbf{\bibinfo{volume}{158}},
  \bibinfo{pages}{165 } (\bibinfo{year}{1985}).

\bibitem[{\citenamefont{Gersten}(2007)}]{GerstenPlasmonics07}
\bibinfo{author}{\bibfnamefont{J.~I.} \bibnamefont{Gersten}},
  \bibinfo{journal}{Plasmonics} \textbf{\bibinfo{volume}{2}},
  \bibinfo{pages}{65} (\bibinfo{year}{2007}).

\bibitem[{\citenamefont{Gersten et~al.}(1979)\citenamefont{Gersten, Birke, and
  Lombardi}}]{GerstenBirkeLombardiPRL79}
\bibinfo{author}{\bibfnamefont{J.~I.} \bibnamefont{Gersten}},
  \bibinfo{author}{\bibfnamefont{R.~L.} \bibnamefont{Birke}}, \bibnamefont{and}
  \bibinfo{author}{\bibfnamefont{J.~R.} \bibnamefont{Lombardi}},
  \bibinfo{journal}{Phys. Rev. Lett.} \textbf{\bibinfo{volume}{43}},
  \bibinfo{pages}{147} (\bibinfo{year}{1979}).

\bibitem[{\citenamefont{Otto et~al.}(1992)\citenamefont{Otto, Mrozek, Grabhorn,
  and Akemann}}]{OttoJPCM92}
\bibinfo{author}{\bibfnamefont{A.}~\bibnamefont{Otto}},
  \bibinfo{author}{\bibfnamefont{I.}~\bibnamefont{Mrozek}},
  \bibinfo{author}{\bibfnamefont{H.}~\bibnamefont{Grabhorn}}, \bibnamefont{and}
  \bibinfo{author}{\bibfnamefont{W.}~\bibnamefont{Akemann}},
  \bibinfo{journal}{Journal of Physics: Condensed Matter}
  \textbf{\bibinfo{volume}{4}}, \bibinfo{pages}{1143} (\bibinfo{year}{1992}).

\bibitem[{\citenamefont{Otto and Futamata}(2006)}]{OttoFutamata}
\bibinfo{author}{\bibfnamefont{A.}~\bibnamefont{Otto}} \bibnamefont{and}
  \bibinfo{author}{\bibfnamefont{M.}~\bibnamefont{Futamata}},
  \emph{\bibinfo{title}{Surface-Enhanced Raman Scattering: Physics and
  Applications}} (\bibinfo{publisher}{Springer}, \bibinfo{year}{2006}), vol.
  \bibinfo{volume}{103}, chap. \bibinfo{chapter}{Electronic Mechanism of SERS},
  pp. \bibinfo{pages}{143--182}.

\bibitem[{\citenamefont{Morton et~al.}(2011)\citenamefont{Morton, Silverstein,
  and Jensen}}]{JensenCR11}
\bibinfo{author}{\bibfnamefont{S.~M.} \bibnamefont{Morton}},
  \bibinfo{author}{\bibfnamefont{D.~W.} \bibnamefont{Silverstein}},
  \bibnamefont{and} \bibinfo{author}{\bibfnamefont{L.}~\bibnamefont{Jensen}},
  \bibinfo{journal}{Chemical Reviews} \textbf{\bibinfo{volume}{111}},
  \bibinfo{pages}{3962} (\bibinfo{year}{2011}).

\bibitem[{\citenamefont{Wu et~al.}(2008)\citenamefont{Wu, Nazin, and
  Ho}}]{HoPRB08}
\bibinfo{author}{\bibfnamefont{S.~W.} \bibnamefont{Wu}},
  \bibinfo{author}{\bibfnamefont{G.~V.} \bibnamefont{Nazin}}, \bibnamefont{and}
  \bibinfo{author}{\bibfnamefont{W.}~\bibnamefont{Ho}}, \bibinfo{journal}{Phys.
  Rev. B} \textbf{\bibinfo{volume}{77}}, \bibinfo{pages}{205430}
  (\bibinfo{year}{2008}).

\bibitem[{\citenamefont{Ioffe et~al.}(2008)\citenamefont{Ioffe, Shamai, Ophir,
  Noy, Yutsis, Kfir, Cheshnovsky, and Selzer}}]{CheshnovskySelzerNatNano08}
\bibinfo{author}{\bibfnamefont{Z.}~\bibnamefont{Ioffe}},
  \bibinfo{author}{\bibfnamefont{T.}~\bibnamefont{Shamai}},
  \bibinfo{author}{\bibfnamefont{A.}~\bibnamefont{Ophir}},
  \bibinfo{author}{\bibfnamefont{G.}~\bibnamefont{Noy}},
  \bibinfo{author}{\bibfnamefont{I.}~\bibnamefont{Yutsis}},
  \bibinfo{author}{\bibfnamefont{K.}~\bibnamefont{Kfir}},
  \bibinfo{author}{\bibfnamefont{O.}~\bibnamefont{Cheshnovsky}},
  \bibnamefont{and} \bibinfo{author}{\bibfnamefont{Y.}~\bibnamefont{Selzer}},
  \bibinfo{journal}{Nature Nanotechnology} \textbf{\bibinfo{volume}{3}},
  \bibinfo{pages}{727} (\bibinfo{year}{2008}).

\bibitem[{\citenamefont{Ward et~al.}(2008)\citenamefont{Ward, Halas, Ciszek,
  Tour, Wu, Nordlander, and Natelson}}]{NatelsonNL08}
\bibinfo{author}{\bibfnamefont{D.~R.} \bibnamefont{Ward}},
  \bibinfo{author}{\bibfnamefont{N.~J.} \bibnamefont{Halas}},
  \bibinfo{author}{\bibfnamefont{J.~W.} \bibnamefont{Ciszek}},
  \bibinfo{author}{\bibfnamefont{J.~M.} \bibnamefont{Tour}},
  \bibinfo{author}{\bibfnamefont{Y.}~\bibnamefont{Wu}},
  \bibinfo{author}{\bibfnamefont{P.}~\bibnamefont{Nordlander}},
  \bibnamefont{and} \bibinfo{author}{\bibfnamefont{D.}~\bibnamefont{Natelson}},
  \bibinfo{journal}{Nano Lett.} \textbf{\bibinfo{volume}{8}},
  \bibinfo{pages}{919} (\bibinfo{year}{2008}).

\bibitem[{\citenamefont{Liu et~al.}(2011)\citenamefont{Liu, Ding, Chen, Wang,
  Tian, Anema, Zhou, Wu, Mao, Xu et~al.}}]{TianNatComm11}
\bibinfo{author}{\bibfnamefont{Z.}~\bibnamefont{Liu}},
  \bibinfo{author}{\bibfnamefont{S.-Y.} \bibnamefont{Ding}},
  \bibinfo{author}{\bibfnamefont{Z.-B.} \bibnamefont{Chen}},
  \bibinfo{author}{\bibfnamefont{X.}~\bibnamefont{Wang}},
  \bibinfo{author}{\bibfnamefont{J.-H.} \bibnamefont{Tian}},
  \bibinfo{author}{\bibfnamefont{J.~R.} \bibnamefont{Anema}},
  \bibinfo{author}{\bibfnamefont{X.-S.} \bibnamefont{Zhou}},
  \bibinfo{author}{\bibfnamefont{D.-Y.} \bibnamefont{Wu}},
  \bibinfo{author}{\bibfnamefont{B.-W.} \bibnamefont{Mao}},
  \bibinfo{author}{\bibfnamefont{X.}~\bibnamefont{Xu}}, \bibnamefont{et~al.},
  \bibinfo{journal}{Nat Commun} \textbf{\bibinfo{volume}{2}},
  \bibinfo{pages}{305} (\bibinfo{year}{2011}).

\bibitem[{\citenamefont{Shamai and Selzer}(2011)}]{SelzerChemSocRev11}
\bibinfo{author}{\bibfnamefont{T.}~\bibnamefont{Shamai}} \bibnamefont{and}
  \bibinfo{author}{\bibfnamefont{Y.}~\bibnamefont{Selzer}},
  \bibinfo{journal}{Chem. Soc. Rev.} \textbf{\bibinfo{volume}{40}},
  \bibinfo{pages}{2293} (\bibinfo{year}{2011}).

\bibitem[{\citenamefont{Ward et~al.}(2011)\citenamefont{Ward, Corley, Tour, and
  Natelson}}]{NatelsonNatNano11}
\bibinfo{author}{\bibfnamefont{D.~R.} \bibnamefont{Ward}},
  \bibinfo{author}{\bibfnamefont{D.~A.} \bibnamefont{Corley}},
  \bibinfo{author}{\bibfnamefont{J.~M.} \bibnamefont{Tour}}, \bibnamefont{and}
  \bibinfo{author}{\bibfnamefont{D.}~\bibnamefont{Natelson}},
  \bibinfo{journal}{Nature Nanotechnology} \textbf{\bibinfo{volume}{6}},
  \bibinfo{pages}{33} (\bibinfo{year}{2011}).

\bibitem[{\citenamefont{Konishi et~al.}(2013)\citenamefont{Konishi, Kiguchi,
  Takase, Nagasawa, Nabika, Ikeda, Uosaki, Ueno, Misawa, and
  Murakoshi}}]{KiguchiMurakoshiJACS13}
\bibinfo{author}{\bibfnamefont{T.}~\bibnamefont{Konishi}},
  \bibinfo{author}{\bibfnamefont{M.}~\bibnamefont{Kiguchi}},
  \bibinfo{author}{\bibfnamefont{M.}~\bibnamefont{Takase}},
  \bibinfo{author}{\bibfnamefont{F.}~\bibnamefont{Nagasawa}},
  \bibinfo{author}{\bibfnamefont{H.}~\bibnamefont{Nabika}},
  \bibinfo{author}{\bibfnamefont{K.}~\bibnamefont{Ikeda}},
  \bibinfo{author}{\bibfnamefont{K.}~\bibnamefont{Uosaki}},
  \bibinfo{author}{\bibfnamefont{K.}~\bibnamefont{Ueno}},
  \bibinfo{author}{\bibfnamefont{H.}~\bibnamefont{Misawa}}, \bibnamefont{and}
  \bibinfo{author}{\bibfnamefont{K.}~\bibnamefont{Murakoshi}},
  \bibinfo{journal}{Journal of the American Chemical Society}
  \textbf{\bibinfo{volume}{135}}, \bibinfo{pages}{1009} (\bibinfo{year}{2013}).

\bibitem[{\citenamefont{Natelson et~al.}(2013)\citenamefont{Natelson, Li, and
  Herzog}}]{NatelsonPCCP13}
\bibinfo{author}{\bibfnamefont{D.}~\bibnamefont{Natelson}},
  \bibinfo{author}{\bibfnamefont{Y.}~\bibnamefont{Li}}, \bibnamefont{and}
  \bibinfo{author}{\bibfnamefont{J.~B.} \bibnamefont{Herzog}},
  \bibinfo{journal}{Phys. Chem. Chem. Phys.} \textbf{\bibinfo{volume}{15}},
  \bibinfo{pages}{5262} (\bibinfo{year}{2013}).

\bibitem[{\citenamefont{Galperin and Nitzan}(2012)}]{MGANPCCP12}
\bibinfo{author}{\bibfnamefont{M.}~\bibnamefont{Galperin}} \bibnamefont{and}
  \bibinfo{author}{\bibfnamefont{A.}~\bibnamefont{Nitzan}},
  \bibinfo{journal}{Phys. Chem. Chem. Phys.} \textbf{\bibinfo{volume}{14}},
  \bibinfo{pages}{9421} (\bibinfo{year}{2012}).

\bibitem[{\citenamefont{Galperin
  et~al.}(2009{\natexlab{a}})\citenamefont{Galperin, Ratner, and
  Nitzan}}]{GalperinRatnerNitzanNL09}
\bibinfo{author}{\bibfnamefont{M.}~\bibnamefont{Galperin}},
  \bibinfo{author}{\bibfnamefont{M.~A.} \bibnamefont{Ratner}},
  \bibnamefont{and} \bibinfo{author}{\bibfnamefont{A.}~\bibnamefont{Nitzan}},
  \bibinfo{journal}{Nano Lett.} \textbf{\bibinfo{volume}{9}},
  \bibinfo{pages}{758} (\bibinfo{year}{2009}{\natexlab{a}}).

\bibitem[{\citenamefont{Galperin
  et~al.}(2009{\natexlab{b}})\citenamefont{Galperin, Ratner, and
  Nitzan}}]{GalperinRatnerNitzanJCP09}
\bibinfo{author}{\bibfnamefont{M.}~\bibnamefont{Galperin}},
  \bibinfo{author}{\bibfnamefont{M.~A.} \bibnamefont{Ratner}},
  \bibnamefont{and} \bibinfo{author}{\bibfnamefont{A.}~\bibnamefont{Nitzan}},
  \bibinfo{journal}{J. Chem. Phys.} \textbf{\bibinfo{volume}{130}},
  \bibinfo{pages}{144109} (\bibinfo{year}{2009}{\natexlab{b}}).

\bibitem[{\citenamefont{Park and Galperin}(2011{\natexlab{a}})}]{ParkMGEPL11}
\bibinfo{author}{\bibfnamefont{T.-H.} \bibnamefont{Park}} \bibnamefont{and}
  \bibinfo{author}{\bibfnamefont{M.}~\bibnamefont{Galperin}},
  \bibinfo{journal}{Europhys. Lett.} \textbf{\bibinfo{volume}{95}},
  \bibinfo{pages}{27001} (\bibinfo{year}{2011}{\natexlab{a}}).

\bibitem[{\citenamefont{Park and Galperin}(2011{\natexlab{b}})}]{ParkMGPRB11}
\bibinfo{author}{\bibfnamefont{T.-H.} \bibnamefont{Park}} \bibnamefont{and}
  \bibinfo{author}{\bibfnamefont{M.}~\bibnamefont{Galperin}},
  \bibinfo{journal}{Phys. Rev. B} \textbf{\bibinfo{volume}{84}},
  \bibinfo{pages}{075447} (\bibinfo{year}{2011}{\natexlab{b}}).

\bibitem[{\citenamefont{Mirjani et~al.}(2012)\citenamefont{Mirjani, Thijssen,
  and Ratner}}]{RatnerJPCC13}
\bibinfo{author}{\bibfnamefont{F.}~\bibnamefont{Mirjani}},
  \bibinfo{author}{\bibfnamefont{J.~M.} \bibnamefont{Thijssen}},
  \bibnamefont{and} \bibinfo{author}{\bibfnamefont{M.~A.}
  \bibnamefont{Ratner}}, \bibinfo{journal}{The Journal of Physical Chemistry C}
  \textbf{\bibinfo{volume}{116}}, \bibinfo{pages}{23120}
  (\bibinfo{year}{2012}).

\bibitem[{\citenamefont{Oren et~al.}(2012)\citenamefont{Oren, Galperin, and
  Nitzan}}]{OrenMGANPRB12}
\bibinfo{author}{\bibfnamefont{M.}~\bibnamefont{Oren}},
  \bibinfo{author}{\bibfnamefont{M.}~\bibnamefont{Galperin}}, \bibnamefont{and}
  \bibinfo{author}{\bibfnamefont{A.}~\bibnamefont{Nitzan}},
  \bibinfo{journal}{Phys. Rev. B} \textbf{\bibinfo{volume}{85}},
  \bibinfo{pages}{115435} (\bibinfo{year}{2012}).

\bibitem[{\citenamefont{Park and Galperin}(2012)}]{ParkMGPST12}
\bibinfo{author}{\bibfnamefont{T.-H.} \bibnamefont{Park}} \bibnamefont{and}
  \bibinfo{author}{\bibfnamefont{M.}~\bibnamefont{Galperin}},
  \bibinfo{journal}{Phys. Scr. T} \textbf{\bibinfo{volume}{151}},
  \bibinfo{pages}{014038} (\bibinfo{year}{2012}).

\bibitem[{\citenamefont{Kaasbjerg et~al.}(2013)\citenamefont{Kaasbjerg,
  Novotn{\' y}, and Nitzan}}]{KaasbjergNitzanPRB13}
\bibinfo{author}{\bibfnamefont{K.}~\bibnamefont{Kaasbjerg}},
  \bibinfo{author}{\bibfnamefont{T.}~\bibnamefont{Novotn{\' y}}},
  \bibnamefont{and} \bibinfo{author}{\bibfnamefont{A.}~\bibnamefont{Nitzan}},
  \bibinfo{journal}{Phys. Rev. B} \textbf{\bibinfo{volume}{88}},
  \bibinfo{pages}{201405} (\bibinfo{year}{2013}).

\bibitem[{\citenamefont{White et~al.}(2014)\citenamefont{White, Tretiak, and
  Galperin}}]{WhiteTretiakNL14}
\bibinfo{author}{\bibfnamefont{A.~J.} \bibnamefont{White}},
  \bibinfo{author}{\bibfnamefont{S.}~\bibnamefont{Tretiak}}, \bibnamefont{and}
  \bibinfo{author}{\bibfnamefont{M.}~\bibnamefont{Galperin}},
  \bibinfo{journal}{Nano Letters} \textbf{\bibinfo{volume}{14}},
  \bibinfo{pages}{699} (\bibinfo{year}{2014}).

\bibitem[{\citenamefont{Burstein et~al.}(1979)\citenamefont{Burstein, Chen,
  Chen, Lundquist, and Tosatti}}]{BursteinSolStCommun79}
\bibinfo{author}{\bibfnamefont{E.}~\bibnamefont{Burstein}},
  \bibinfo{author}{\bibfnamefont{Y.}~\bibnamefont{Chen}},
  \bibinfo{author}{\bibfnamefont{C.}~\bibnamefont{Chen}},
  \bibinfo{author}{\bibfnamefont{S.}~\bibnamefont{Lundquist}},
  \bibnamefont{and} \bibinfo{author}{\bibfnamefont{E.}~\bibnamefont{Tosatti}},
  \bibinfo{journal}{Solid State Communications} \textbf{\bibinfo{volume}{29}},
  \bibinfo{pages}{567 } (\bibinfo{year}{1979}).

\bibitem[{\citenamefont{Akemann and Otto}(1994)}]{AkemannOttoSurfSci94}
\bibinfo{author}{\bibfnamefont{W.}~\bibnamefont{Akemann}} \bibnamefont{and}
  \bibinfo{author}{\bibfnamefont{A.}~\bibnamefont{Otto}},
  \bibinfo{journal}{Surface Science} \textbf{\bibinfo{volume}{307-309, Part
  B}}, \bibinfo{pages}{1071} (\bibinfo{year}{1994}).

\bibitem[{\citenamefont{Itoh et~al.}(2006)\citenamefont{Itoh, Biju, Ishikawa,
  Kikkawa, Hashimoto, Ikehata, and Ozaki}}]{ItohJCP06}
\bibinfo{author}{\bibfnamefont{T.}~\bibnamefont{Itoh}},
  \bibinfo{author}{\bibfnamefont{V.}~\bibnamefont{Biju}},
  \bibinfo{author}{\bibfnamefont{M.}~\bibnamefont{Ishikawa}},
  \bibinfo{author}{\bibfnamefont{Y.}~\bibnamefont{Kikkawa}},
  \bibinfo{author}{\bibfnamefont{K.}~\bibnamefont{Hashimoto}},
  \bibinfo{author}{\bibfnamefont{A.}~\bibnamefont{Ikehata}}, \bibnamefont{and}
  \bibinfo{author}{\bibfnamefont{Y.}~\bibnamefont{Ozaki}},
  \bibinfo{journal}{The Journal of Chemical Physics}
  \textbf{\bibinfo{volume}{124}}, \bibinfo{eid}{134708} (\bibinfo{year}{2006}).

\bibitem[{\citenamefont{Marhaba et~al.}(2009)\citenamefont{Marhaba, Bachelier,
  Bonnet, Broyer, Cottancin, Grillet, Lermé, Vialle, and
  Pellarin}}]{MarhabaJPCC09}
\bibinfo{author}{\bibfnamefont{S.}~\bibnamefont{Marhaba}},
  \bibinfo{author}{\bibfnamefont{G.}~\bibnamefont{Bachelier}},
  \bibinfo{author}{\bibfnamefont{C.}~\bibnamefont{Bonnet}},
  \bibinfo{author}{\bibfnamefont{M.}~\bibnamefont{Broyer}},
  \bibinfo{author}{\bibfnamefont{E.}~\bibnamefont{Cottancin}},
  \bibinfo{author}{\bibfnamefont{N.}~\bibnamefont{Grillet}},
  \bibinfo{author}{\bibfnamefont{J.}~\bibnamefont{Lermé}},
  \bibinfo{author}{\bibfnamefont{J.-L.} \bibnamefont{Vialle}},
  \bibnamefont{and} \bibinfo{author}{\bibfnamefont{M.}~\bibnamefont{Pellarin}},
  \bibinfo{journal}{The Journal of Physical Chemistry C}
  \textbf{\bibinfo{volume}{113}}, \bibinfo{pages}{4349} (\bibinfo{year}{2009}).

\bibitem[{\citenamefont{Yampolsky et~al.}(2014)\citenamefont{Yampolsky,
  Fishman, Dey, Hulkko, Banik, Potma, and Apkarian}}]{ApkarianNatPhoton14}
\bibinfo{author}{\bibfnamefont{S.}~\bibnamefont{Yampolsky}},
  \bibinfo{author}{\bibfnamefont{D.~A.} \bibnamefont{Fishman}},
  \bibinfo{author}{\bibfnamefont{S.}~\bibnamefont{Dey}},
  \bibinfo{author}{\bibfnamefont{E.}~\bibnamefont{Hulkko}},
  \bibinfo{author}{\bibfnamefont{M.}~\bibnamefont{Banik}},
  \bibinfo{author}{\bibfnamefont{E.~O.} \bibnamefont{Potma}}, \bibnamefont{and}
  \bibinfo{author}{\bibfnamefont{V.~A.} \bibnamefont{Apkarian}},
  \bibinfo{journal}{Nat. Photon.} \textbf{\bibinfo{volume}{8}},
  \bibinfo{pages}{650} (\bibinfo{year}{2014}).

\bibitem[{\citenamefont{Ganichev et~al.}(1998)\citenamefont{Ganichev, Ziemann,
  Gleim, Prettl, Yassievich, Perel, Wilke, and Haller}}]{GanichevPRL98}
\bibinfo{author}{\bibfnamefont{S.~D.} \bibnamefont{Ganichev}},
  \bibinfo{author}{\bibfnamefont{E.}~\bibnamefont{Ziemann}},
  \bibinfo{author}{\bibfnamefont{T.}~\bibnamefont{Gleim}},
  \bibinfo{author}{\bibfnamefont{W.}~\bibnamefont{Prettl}},
  \bibinfo{author}{\bibfnamefont{I.~N.} \bibnamefont{Yassievich}},
  \bibinfo{author}{\bibfnamefont{V.~I.} \bibnamefont{Perel}},
  \bibinfo{author}{\bibfnamefont{I.}~\bibnamefont{Wilke}}, \bibnamefont{and}
  \bibinfo{author}{\bibfnamefont{E.~E.} \bibnamefont{Haller}},
  \bibinfo{journal}{Phys. Rev. Lett.} \textbf{\bibinfo{volume}{80}},
  \bibinfo{pages}{2409} (\bibinfo{year}{1998}).

\bibitem[{\citenamefont{Ward et~al.}(2010)\citenamefont{Ward, Huser, Pauly,
  Cuevas, and Natelson}}]{CuevasNatelsonNN10}
\bibinfo{author}{\bibfnamefont{D.~R.} \bibnamefont{Ward}},
  \bibinfo{author}{\bibfnamefont{F.}~\bibnamefont{Huser}},
  \bibinfo{author}{\bibfnamefont{F.}~\bibnamefont{Pauly}},
  \bibinfo{author}{\bibfnamefont{J.~C.} \bibnamefont{Cuevas}},
  \bibnamefont{and} \bibinfo{author}{\bibfnamefont{D.}~\bibnamefont{Natelson}},
  \bibinfo{journal}{Nature Nano} \textbf{\bibinfo{volume}{5}},
  \bibinfo{pages}{732} (\bibinfo{year}{2010}).

\bibitem[{\citenamefont{Vadai et~al.}(2013)\citenamefont{Vadai, Nachman,
  Ben-Zion, Bürkle, Pauly, Cuevas, and Selzer}}]{SelzerJPCL13}
\bibinfo{author}{\bibfnamefont{M.}~\bibnamefont{Vadai}},
  \bibinfo{author}{\bibfnamefont{N.}~\bibnamefont{Nachman}},
  \bibinfo{author}{\bibfnamefont{M.}~\bibnamefont{Ben-Zion}},
  \bibinfo{author}{\bibfnamefont{M.}~\bibnamefont{Bürkle}},
  \bibinfo{author}{\bibfnamefont{F.}~\bibnamefont{Pauly}},
  \bibinfo{author}{\bibfnamefont{J.~C.} \bibnamefont{Cuevas}},
  \bibnamefont{and} \bibinfo{author}{\bibfnamefont{Y.}~\bibnamefont{Selzer}},
  \bibinfo{journal}{The Journal of Physical Chemistry Letters}
  \textbf{\bibinfo{volume}{4}}, \bibinfo{pages}{2811} (\bibinfo{year}{2013}).

\bibitem[{\citenamefont{Galperin and Nitzan}(2011{\natexlab{a}})}]{MGANJPCL11}
\bibinfo{author}{\bibfnamefont{M.}~\bibnamefont{Galperin}} \bibnamefont{and}
  \bibinfo{author}{\bibfnamefont{A.}~\bibnamefont{Nitzan}},
  \bibinfo{journal}{J. Phys. Chem. Lett.} \textbf{\bibinfo{volume}{2}},
  \bibinfo{pages}{2110} (\bibinfo{year}{2011}{\natexlab{a}}).

\bibitem[{\citenamefont{Galperin and Nitzan}(2011{\natexlab{b}})}]{MGANPRB11}
\bibinfo{author}{\bibfnamefont{M.}~\bibnamefont{Galperin}} \bibnamefont{and}
  \bibinfo{author}{\bibfnamefont{A.}~\bibnamefont{Nitzan}},
  \bibinfo{journal}{Phys. Rev. B} \textbf{\bibinfo{volume}{84}},
  \bibinfo{pages}{195325} (\bibinfo{year}{2011}{\natexlab{b}}).

\bibitem[{\citenamefont{Banik et~al.}(2012)\citenamefont{Banik, El-Khoury, Nag,
  Rodriguez-Perez, Guarrottxena, Bazan, and Apkarian}}]{ApkarianACSNano12}
\bibinfo{author}{\bibfnamefont{M.}~\bibnamefont{Banik}},
  \bibinfo{author}{\bibfnamefont{P.~Z.} \bibnamefont{El-Khoury}},
  \bibinfo{author}{\bibfnamefont{A.}~\bibnamefont{Nag}},
  \bibinfo{author}{\bibfnamefont{A.}~\bibnamefont{Rodriguez-Perez}},
  \bibinfo{author}{\bibfnamefont{N.}~\bibnamefont{Guarrottxena}},
  \bibinfo{author}{\bibfnamefont{G.~C.} \bibnamefont{Bazan}}, \bibnamefont{and}
  \bibinfo{author}{\bibfnamefont{V.~A.} \bibnamefont{Apkarian}},
  \bibinfo{journal}{ACS Nano} \textbf{\bibinfo{volume}{6}},
  \bibinfo{pages}{10343} (\bibinfo{year}{2012}).

\bibitem[{\citenamefont{Al\`u and Engheta}(2008)}]{EnghetaPRB08}
\bibinfo{author}{\bibfnamefont{A.}~\bibnamefont{Al\`u}} \bibnamefont{and}
  \bibinfo{author}{\bibfnamefont{N.}~\bibnamefont{Engheta}},
  \bibinfo{journal}{Phys. Rev. B} \textbf{\bibinfo{volume}{78}},
  \bibinfo{pages}{195111} (\bibinfo{year}{2008}).

\bibitem[{\citenamefont{Hugall and Baumberg}(2015)}]{HugallBaumbergNL15}
\bibinfo{author}{\bibfnamefont{J.~T.} \bibnamefont{Hugall}} \bibnamefont{and}
  \bibinfo{author}{\bibfnamefont{J.~J.} \bibnamefont{Baumberg}},
  \bibinfo{journal}{Nano Lett.} \textbf{\bibinfo{volume}{15}},
  \bibinfo{pages}{2600} (\bibinfo{year}{2015}).

\bibitem[{\citenamefont{Banik}(2014)}]{BanikPhD14}
\bibinfo{author}{\bibfnamefont{M.}~\bibnamefont{Banik}}, Ph.D. thesis,
  \bibinfo{school}{UC Irvine}, \bibinfo{address}{UC Irvine}
  (\bibinfo{year}{2014}).

\bibitem[{\citenamefont{Galperin
  et~al.}(2015{\natexlab{a}})\citenamefont{Galperin, Ratner, and
  Nitzan}}]{MGRatnerNitzanJCP15}
\bibinfo{author}{\bibfnamefont{M.}~\bibnamefont{Galperin}},
  \bibinfo{author}{\bibfnamefont{M.~A.} \bibnamefont{Ratner}},
  \bibnamefont{and} \bibinfo{author}{\bibfnamefont{A.}~\bibnamefont{Nitzan}},
  \bibinfo{journal}{The Journal of Chemical Physics}
  \textbf{\bibinfo{volume}{142}}, \bibinfo{eid}{137101}
  (\bibinfo{year}{2015}{\natexlab{a}}).

\bibitem[{\citenamefont{Galperin
  et~al.}(2015{\natexlab{b}})\citenamefont{Galperin, Ratner, and
  Nitzan}}]{MGRatnerNitzanARXIV15}
\bibinfo{author}{\bibfnamefont{M.}~\bibnamefont{Galperin}},
  \bibinfo{author}{\bibfnamefont{M.~A.} \bibnamefont{Ratner}},
  \bibnamefont{and} \bibinfo{author}{\bibfnamefont{A.}~\bibnamefont{Nitzan}},
  \emph{\bibinfo{title}{On optical spectroscopy of molecular junctions}},
  \bibinfo{howpublished}{arxiv:1503.03890}
  (\bibinfo{year}{2015}{\natexlab{b}}),
  \urlprefix\url{http://arxiv.org/abs/1503.03890}.

\bibitem[{\citenamefont{Haug and Jauho}(2008)}]{HaugJauho_2008}
\bibinfo{author}{\bibfnamefont{H.}~\bibnamefont{Haug}} \bibnamefont{and}
  \bibinfo{author}{\bibfnamefont{A.-P.} \bibnamefont{Jauho}},
  \emph{\bibinfo{title}{Quantum {K}inetics in {T}ransport and {O}ptics of
  {S}emiconductors}}, vol. \bibinfo{volume}{123}
  (\bibinfo{publisher}{Springer}, \bibinfo{address}{Berlin Heidelberg},
  \bibinfo{year}{2008}).

\end{thebibliography}

\end{document}